\documentclass[10pt,journal,compsoc]{IEEEtran}

\ifCLASSOPTIONcompsoc
  \usepackage[nocompress]{cite}
\else
  \usepackage{cite}
\fi

\ifCLASSINFOpdf
   \usepackage[pdftex]{graphicx}
   \graphicspath{{../pdf/}{../jpeg/}}
   \DeclareGraphicsExtensions{.pdf,.jpeg,.png}
\else
   \usepackage[dvips]{graphicx}
   \graphicspath{{../eps/}}
   \DeclareGraphicsExtensions{.eps}
\fi

\usepackage{graphicx}
\usepackage{comment}
\usepackage{multirow} 
\usepackage[table,xcdraw]{xcolor}
\usepackage{caption}
\usepackage{subcaption}
\usepackage{epstopdf}
\usepackage{amsmath}
\usepackage{array}
\usepackage{tikz}
\usepackage[export]{adjustbox}
\usepackage{hyperref}
\usepackage{multicol} 
\usepackage{booktabs} 
\definecolor{Henan}{RGB}{255,0,0}

\usepackage[table,xcdraw]{xcolor}
\usepackage{wrapfig}
\usepackage{enumitem}
\usepackage{textcomp}
\usepackage{expdlist}
\usepackage[most]{tcolorbox}
\usepackage[utf8]{inputenc}
\usepackage{multicol}       
\usepackage{dblfloatfix}    
\usepackage{xspace}
\usepackage[finalnew]{trackchanges}

\newcommand{\changes}[1]{\textcolor{blue}{#1}}
\renewcommand{\changes}[1]{#1}

\newcommand{\changesr}[1]{\textcolor{blue}{#1}}
\renewcommand{\changesr}[1]{#1}

\newcommand{\Lengthylengthy}{\changes{$Length_ylength_y$}\xspace}
\newcommand{\lengthylengthy}{\changes{$length_ylength_y$}\xspace}
\newcommand{\lyly}{$L_yL_y$\xspace}
\newcommand{\Lengthylengthx}{\changes{$Length_ylength_x$}\xspace}
\newcommand{\lengthylengthx}{\changes{$length_ylength_x$}\xspace}
\newcommand{\lylx}{$L_yL_x$\xspace}
\newcommand{\Lengthycolor}{\changes{$Length_ycolor$}\xspace}
\newcommand{\lengthycolor}{\changes{$length_ycolor$}\xspace}
\newcommand{\lc}{$LC$\xspace}
\newcommand{\Lengthytexture}{\changes{$Length_ytexture$}\xspace}
\newcommand{\lengthytexture}{\changes{$length_ytexture$}\xspace}
\newcommand{\lt}{$LT$\xspace}
\newcommand{\Lengthycolorlengthx}{\changes{$Length_ycolor/length_x$}\xspace}
\newcommand{\lengthycolorlengthx}{\changes{$length_ycolor/length_x$}\xspace}
\newcommand{\lcl}{$LCL$\xspace}

\begin{document}
\title{
\changes{Evaluating Glyph Design for Showing Large-Magnitude-Range Quantum Spins}
}


\author{Henan~Zhao,
        Garnett~W.~Bryant,
        Wesley~Griffin,
        Judith~E.~Terrill,
        Jian Chen
\IEEEcompsocitemizethanks{\IEEEcompsocthanksitem Henan Zhao is with  University of Maryland, Baltimore County. E-mail: henan1@umbc.edu.
\IEEEcompsocthanksitem Garnett W. Bryant and Judith E. Terrill are with the National Institute of Standards and Technology. E-mail: \{garnett.bryant, judith.terrill\}@nist.gov.
\IEEEcompsocthanksitem Wesley Griffin is with Stellar Science. E-mail: griffin5@umbc.edu.
\IEEEcompsocthanksitem Jian Chen is with The Ohio State University. E-mail: chen.8028@osu.edu.
}
}

\markboth{Journal of \LaTeX\ Class Files,~Vol.~14, No.~8, August~2015}%
{Zhao \MakeLowercase{\textit{et al.}}: --}

\IEEEtitleabstractindextext{%
\begin{abstract}
\changesr{We present experimental results to explore a form of bivariate glyphs for representing large-magnitude-range vectors. 
The glyphs meet two conditions: (1) two visual  dimensions are separable; and (2) one of the two visual dimensions uses a categorical representation (e.g., a categorical colormap).
We evaluate how much these
two conditions determine the bivariate glyphs' effectiveness.
The first experiment asks participants to perform three 
local tasks requiring reading no more than
two glyphs. 
The second experiment scales up the search space in global tasks when participants must look at the entire scene of hundreds of vector glyphs to get an answer.
Our results support that 
the first condition  is necessary for local tasks when a few items are compared. But it is not enough for understanding a large amount of data. The second condition is necessary for perceiving global structures of examining very complex datasets.
Participants' comments reveal that the categorical features in the bivariate glyphs trigger emergent optimal viewers' behaviors. 
This work contributes to perceptually accurate glyph representations for revealing patterns from large scientific results.
We release source code, quantum physics data, training documents, participants' answers, and statistical analyses for reproducible science at \href{https://osf.io/4xcf5/?view_only=94123139df9c4ac984a1e0df811cd580}{$https://osf.io/4xcf5/?view_only=94123139df9c4ac984a1e0df811cd580$}.}
\end{abstract}

}

\maketitle

\IEEEdisplaynontitleabstractindextext

\IEEEpeerreviewmaketitle

\IEEEraisesectionheading{\section{Introduction}\label{sec:introduction}}



\IEEEPARstart{B}{ivarate}
glyph  visualization is a common form of visual design in  which a dataset is depicted by  two  visual variables, often   
chosen   from  a  set  of  perceptually  independent graphical dimensions of shape, color, texture, size, orientation, curvature, and  so on{~\cite{fuchs2017systematic,ware2009quantitative}}. 
A bivariate glyph design~\cite{henan2017} has  been  broadly applied to reveal atom spin behaviors for quantum physicists at National Institute of Standards and Technology
(NIST) 
to examine experimental results; thanks to their team's Nobel-prize-winning 
simulations~\cite{wineland2013nobel}. 
\changes{Quantum physicists world-wide can now manipulate many individual 
quantum systems to study complex atom and sub-atom interactions.
\changesr{Because atoms can be in multiple states simultaneously 
}
and because these spin magnitudes 
are large in range and often vary greatly in local regions,
computational solutions still do not exist to characterize the spin behaviors. Today's quantum physicists rely on visualization to interpret simulation results.
}

\begin{figure*}[!t]
  \centering
\includegraphics[width=0.85\linewidth]{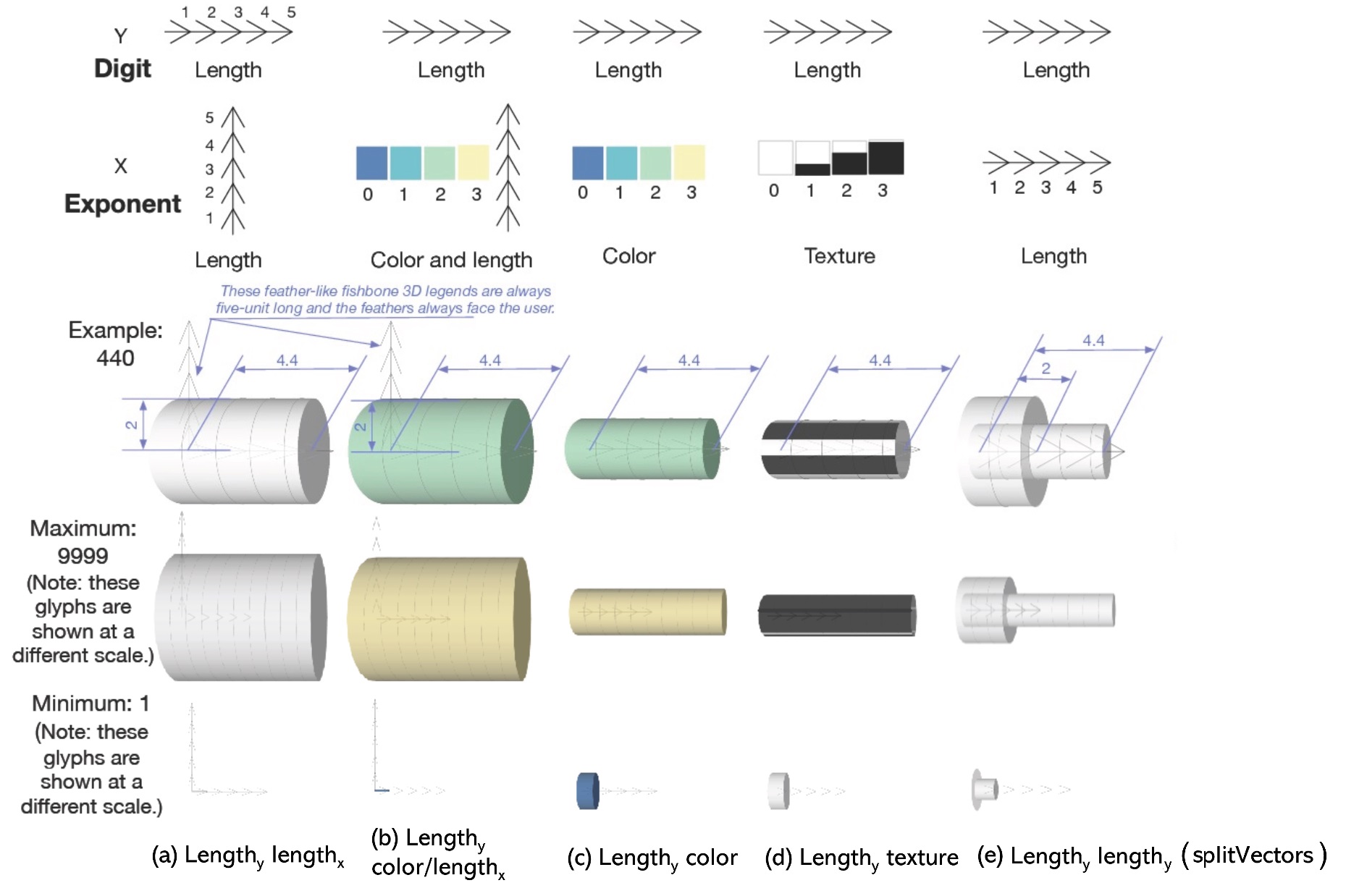}
\caption{
\changesr{Illustration of five bivariate configurations of vector magnitudes $\in(0, 9,999]$. 
Three examples show vector magnitudes $440$ ($4.4 \times 10^2$), $9,999$ ($9.9\times10^3$), and $1$ ($1\times10^0$). Take 440 as an example,
\lengthylengthx (a) maps $4.4$ (digit) and $2$ (exponent) to \textit{lengths} along the y and x axes accordingly ( $length_y\;length_x$); 
(b)-(e) have the same digit-to-$length_y$ representation as (a). The exponent representations are manipulated to be  (1) more integral or separable from $length_y$ and (2) more or less categorical.
(b) \lengthycolorlengthx uses color to double-code exponent compared to (a). The exponents in (c), (d), and (e) use color, texture, or outer cylinder length accordingly. 
Our experimental results support that more separable dimensions lead to more perceptually accurate glyphs. The higher the separability, the higher the accuracy.
Also, using a more categorical feature (e.g., color in (c)) of one of the variables
benefited efficiency and accuracy.}
}
\label{fig:teaser}
\end{figure*}

\begin{figure*}[!tp]
\centering
	\begin{subfigure}[t]{0.33\textwidth}
		\centering
\includegraphics[width=\textwidth]{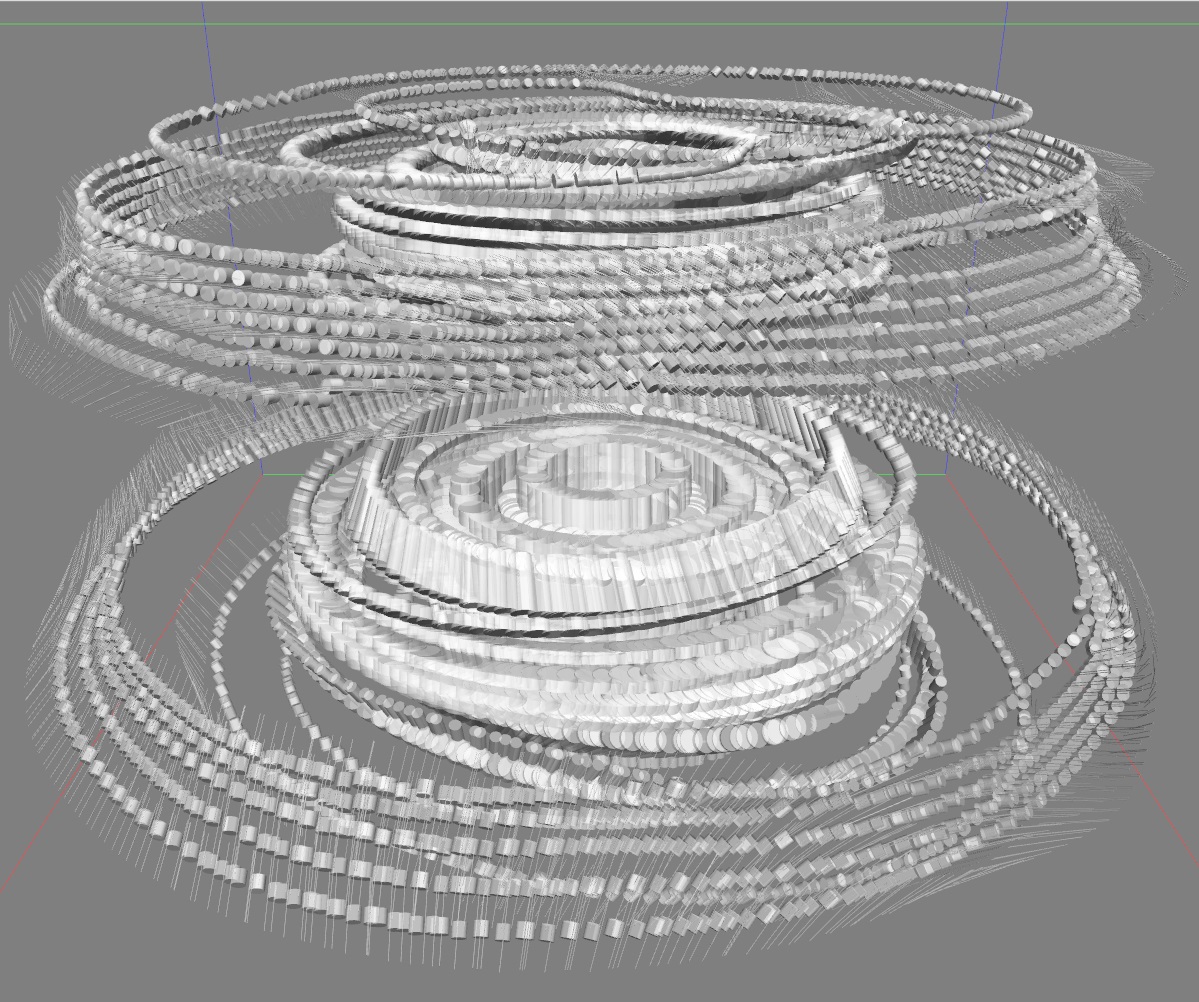}
		\caption{\Lengthylengthx (\lylx) (integral)}
	\end{subfigure}
	\begin{subfigure}[t]{0.33\textwidth}
	\includegraphics[width=\textwidth]{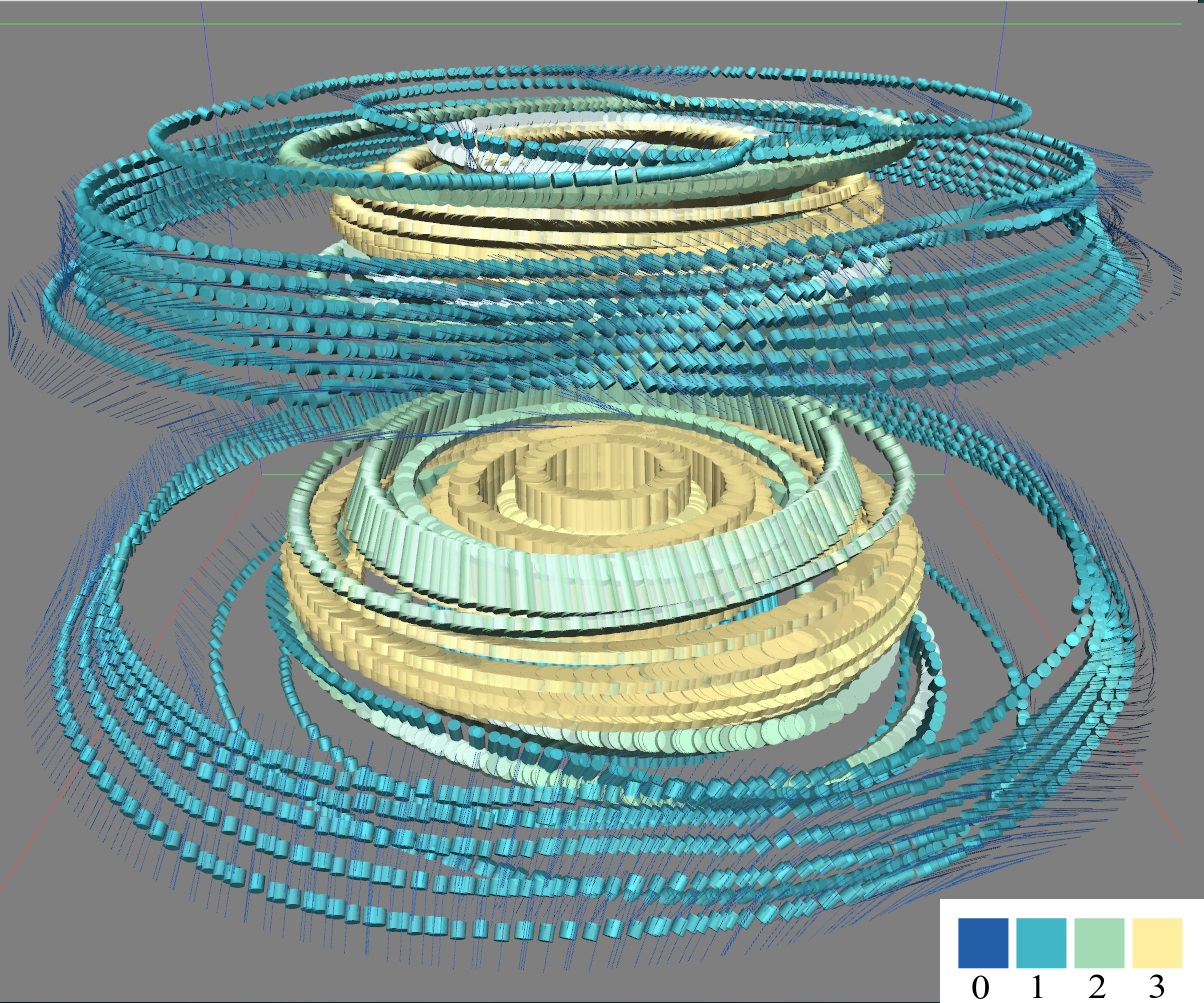}
		\caption{\Lengthycolorlengthx (\lcl) (redundant encoding)}
	\end{subfigure}
	\begin{subfigure}[t]{0.33\textwidth}
		\centering
\includegraphics[width=\textwidth]{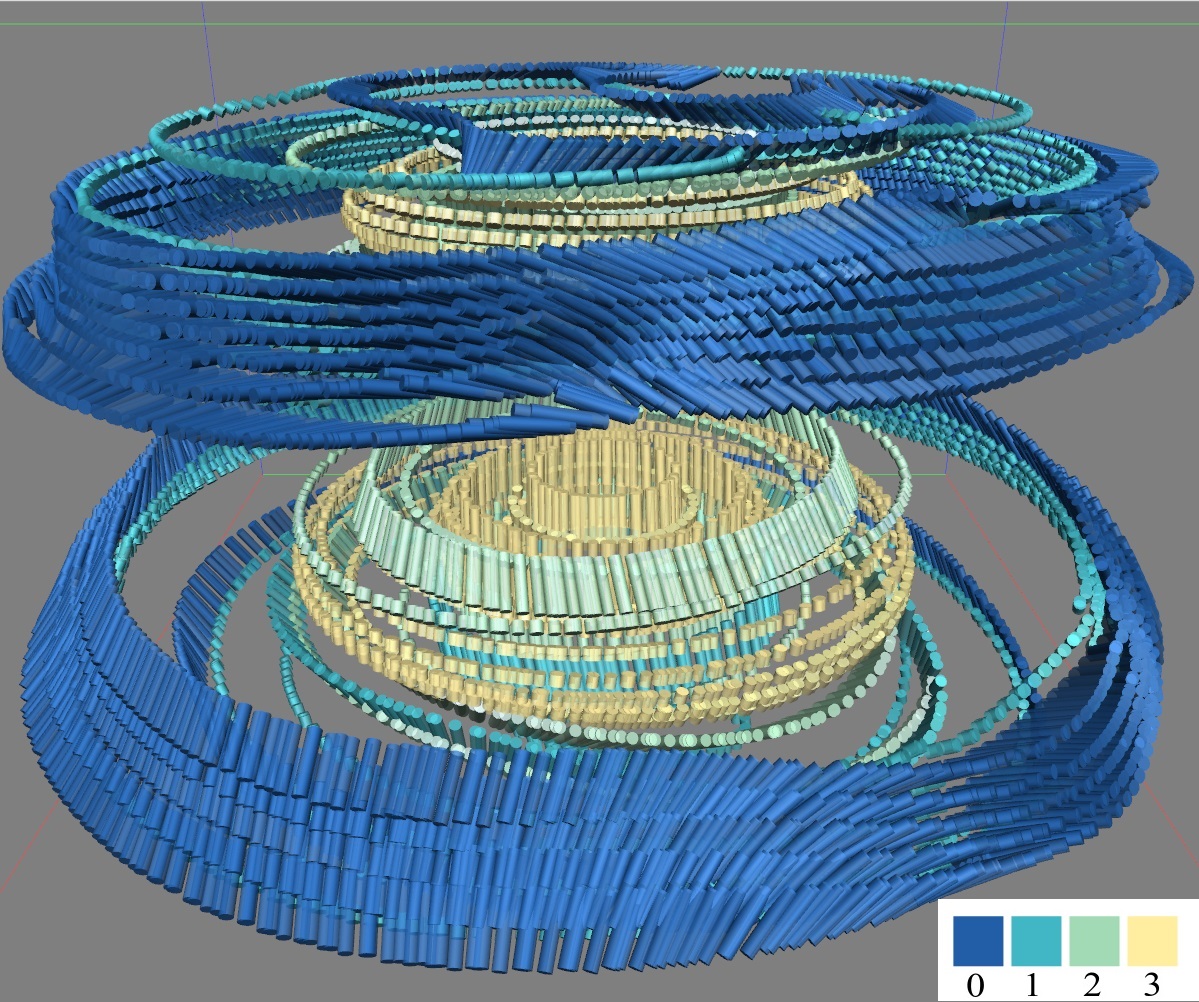}
		\caption{\Lengthycolor (\lc) (separable)}
	\end{subfigure}
    
    	\begin{subfigure}[t]{0.33\textwidth}
		\centering
\includegraphics[width=\textwidth]{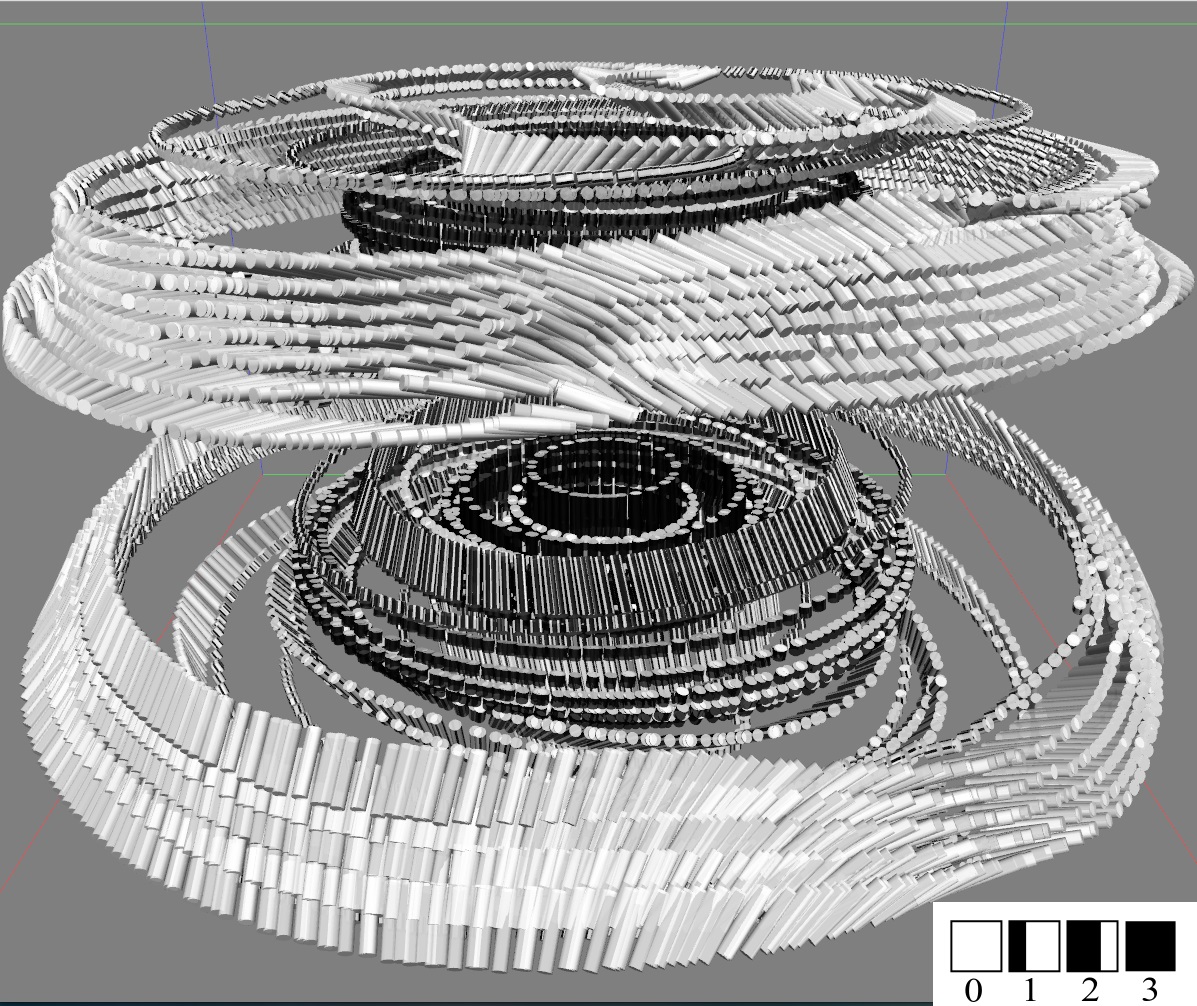}
		\caption{\Lengthytexture (\lt) (separable)}
	\end{subfigure}
    	\begin{subfigure}[t]{0.33\textwidth}
		\centering
\includegraphics[width=\textwidth]{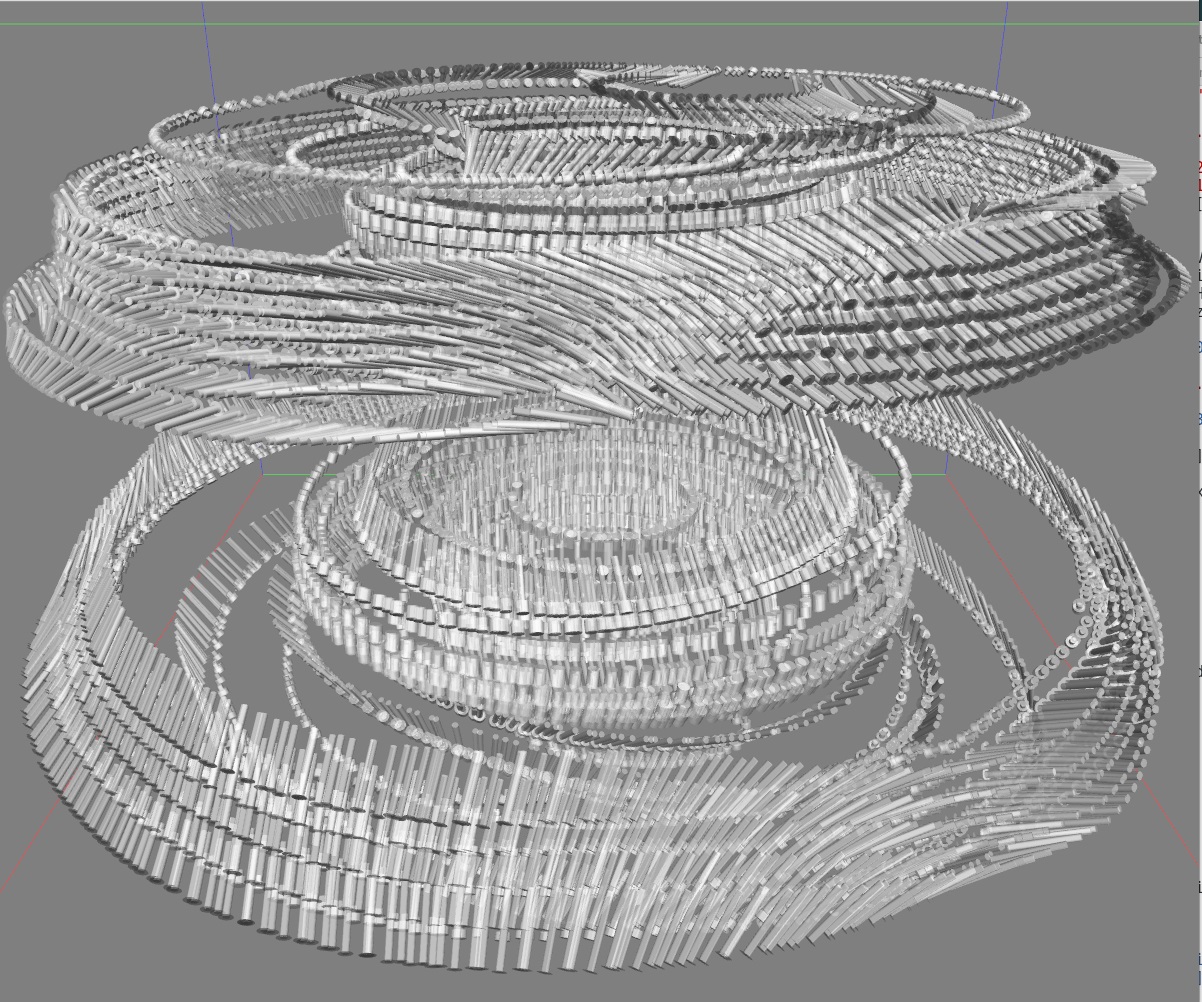}
		\caption{\Lengthylengthy (\textit{splitVectors}, \lyly) \cite{henan2017}} 
\end{subfigure}
	\begin{subfigure}[t]{0.318\textwidth}    
	\includegraphics[width=\textwidth]{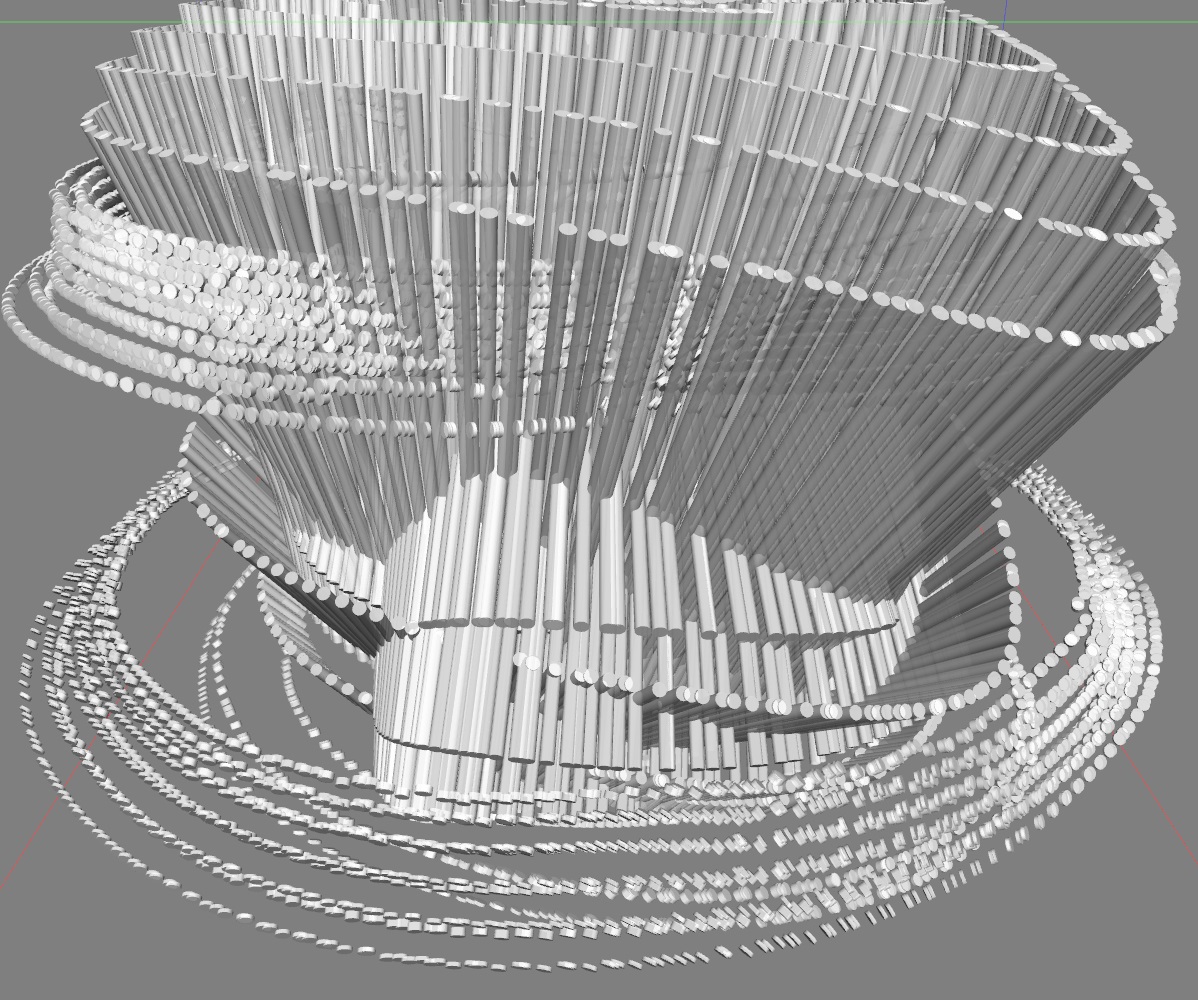}
		\caption{\textit{Linear}}
	\end{subfigure}
\caption{Real-world large-magnitude-range quantum physics simulation results shown using (a)-(e) five bivariate feature-pairs and (f) a traditional linear representation. \lc, \lcl, and \lt can reveal scene spatial structures. 
\changesr{We anticipate that two conditions determine the glyph efficiency: (1) the bivariate glyph uses two separable dimensions; and (2) one of the two dimensions uses a categorical representation thus can reveal global structures in data. The first condition is necessary for local tasks when a few items are compared. The second condition is needed for inspecting the entire scene.
%
}
}
\label{fig:cases}
\end{figure*}

On  the  visualization side, 
the initial design and evaluation of large-magnitude-range spin vector visualizations use scientific notation to depict digit and exponent as two concentric cylinders~\cite{henan2017}: inside and outside tube-lengths (\lengthylengthy or \lyly or \textit{splitVectors}) are mapped to digit and power of spin magnitude accordingly (Figure{~\ref{fig:teaser}}e).
A three-dimensional (3D) bivariate glyph scene of this \textit{splitVectors} design (Figure{~\ref{fig:cases}}e) achieves up to ten times greater accuracy than the traditional direct approach (\textit{Linear}, Figure{~\ref{fig:cases}}f) for 
reading a vector magnitude of a single spin or deriving ratios
between two spin magnitudes. However, this design also 
increases task completion time for an apparently simple comparison task between two 
magnitudes in three dimensions (3D): the traditional direct approach of \textit{Linear} (Figure{~\ref{fig:cases}f}) is significantly faster
than \textit{splitVectors} \add{(Figure{~\ref{fig:cases}}e)}.

One may frame  this large-magnitude-range issue as a visual design problem: 
\textit{how can we depict a scalar value using bivariate visual features to help quantum physicists examine complex spatial data?}
Intuitively,
\changes{
since all tasks in previous study involve a single or at most two spin locations, 
human visual system would integrate the two component parts (digit and exponent terms) of a quantum spin into one gestalt
before comparing the results~\cite{treisman1980feature}.
\changesr{Since relating the digit and exponent to the two \textit{size} features demands a focused attention mode of visual processing, 
a viewer would take longer
to process two component parts in \textit{splitVectors} compared to a single linear mapping.
}
We term this explanation the \textit{object-level hypothesis} where a viewer responds 
to combine two component parts of a value represented in a glyph to its
original 
scalar value (here the magnitude).
}

\changes{However, the \textit{object-level} processing may be neither efficient nor necessary.
} 
\changesr{For example, Borgo et al.~\cite{borgo2013glyph} state that ``... \textit{effective glyph design should encompass a non-conflicting set of separable retinal variables}''.}
Now, for our examples, if we increase the bivariate feature separability by replacing the exponent-to-length mapping in Figures{~\ref{fig:teaser}}e and{~\ref{fig:cases}}e to the exponent-to-color mapping in Figures{~\ref{fig:teaser}}c and{~\ref{fig:cases}}c for comparison tasks, it would be counterproductive for our attention first to visit each glyph to compute the magnitude. Instead, the global categorical color (hue) can guide our attention to first compare the exponent, prior to 
compare vector lengths (digit). 
\changes{
In these cases, 
no object-level attentive processing of bivariate features is needed as long as
\changesr{the two} color hues
can be easily recognizable. 

Further considering the quantum physicists' task relevant to multiple  objects (e.g., find maximum among hundreds of vectors) (Figure{~\ref{fig:cases}}), 
viewers are likely to check the color legend and then use color to first divide the scene into subregions, prior to use length for detailed comparisons within the 
yellow region 
(Figures{~\ref{fig:cases}}b and  ~\ref{fig:cases}c).
The colorful scene context benefits the reduction of search to a much smaller scale via global statistics of the scene.
}
Coincidentally, this first impression of the data to drive structural and statistical information is also called \textit{scene-level processing}~\cite{wolfe2019preattentive}; Wolfe called features guiding this top-down task-driven attention behaviors as \textit{scene} features.  
\changesr{
Scene features are also \textit{preattentive} and can guide attention in visual search toward a target
~\cite{wolfe2021guided}, perhaps due to fast ensemble processing~\cite{ariely2001seeing}.}

\changesr{Taken together, an effective  design of bivariate glyphs is likely to be influenced by two conditions: separable dimensions, with one of them being a pre-attentive scene feature.
These two factors are not necessarily independent. For example,
For the first factor, we can follow Borgo et al.~\cite{borgo2013glyph} and Ware~\cite{ware2020information} for  ``\textit{a non-conflicting set of separable retinal variables}''.} 
To meet the both conditions to choose the scene feature,
we can give preferences of the separable pair when one of the variables is categorical.
This is because categorical features are likely to be better at facilitating the perception of a group of objects in the scene~\cite{chen2021ensemble, sekimoto2022ensemble, wolfe2019preattentive}. 
We in this work compared several separable-integral pairs, \textit{length-color} (Figures~\ref{fig:teaser}b,~\ref{fig:cases}b,~\ref{fig:teaser}c,~\ref{fig:cases}c), \textit{length-texture} (Figures~\ref{fig:teaser}d, ~\ref{fig:cases}d), and 
\textit{length-length} 
(Figures~\ref{fig:teaser}a, ~\ref{fig:cases}a).
Among the three features of color,  texture, and size,
color is categorical and thus ``most recognizable''. Color ensembles are preattentive and permit visual selection at a glance~\cite{Maule14}.
We purposefully select texture patterns by varying
the amount of dark on white, thus introducing luminance variations when many vectors are examined together (Figure{~\ref{fig:cases}}d).
Compared to the continuous random noise in Urness et al.{~\cite{urness2003effectively}}, ours is for
discrete quantities and thus 
uses regular scale variations.
\changesr{
When coupled with separable features,
we hypothesize that \textit{
highly distinguishable
separable dimension pairs, with one being categorical might 
encourage preattentive global processing
to reduce task completion time and be more accurate.} } 

We tested this hypothesis in two experiments with  
six tasks using
four 
pairs to compare against
the \lengthylengthy (separable) in Zhao et al.~\cite{henan2017}: 
\lengthylengthx (integral), \lengthycolor (separable), \lengthytexture (separable), 
and \lengthycolorlengthx (redundant and separable).
Since we predicted that separable dimensions with more preattentive features would reduce the task completion time, 
\lengthycolor and \lengthycolorlengthx
might achieve more efficiency without hampering accuracy than other bivariate pairs.

This work makes the following contributions: 

\begin{itemize}

\item Empirically validates that bivariate-glyphs encoded by highly  separable dimensions would {improve comparison \changesr{task completion time}} (Exp I). 


\item 
Is the first to 
\changesr{evaluate categorical features in bivirate-glyphs to leverage
the benefits of the global scene features}
(Exp II).

\item
Offers a rank order of separable variables for 3D glyph design and 
shows that the separable pairs \lengthycolor and \lengthytexture are among the most effective and efficient {feature pairs.}

\item
\changes{Reveals a novel visual design method for scalable search in big-data.
}

\end{itemize}

\section{Theoretical Foundations in Perception and Vision Sciences}

\changes{At least four perceptual and vision science theories have inspired our work: integral and separable dimensions~\cite{garner1970integrality}, preattentive 
scene features~\cite{wolfe2019preattentive, wolfe2021guided, healey1999large, healey1995visualizing}, 
feature ranking, and monotonicity~\cite{ware2009quantitative}. }

\textbf{Integral and Separable Dimensions.}
Garner and  Felfoldy's  seminal work on integral and separable dimensions~\cite{garner1970integrality} has inspired many visualization design guidelines. 
Ware~\cite{ware2020information} suggests a continuum from more integral to more separable pairs: 
\textit{(red-green)}-\textit{(yellow-blue)}, \textit{$size_x$}-\textit{$size_y$},  \textit{color-shape/size/orientation}, \textit{motion-shape/size/orientation}, 
\textit{motion-color}, and \textit{group position-color}. 
His subsequent award-winning bivariate study~\cite{ware2009quantitative} using \textit{hue-size}, \textit{hue-luminance}, and \textit{hue-texton} (texture) 
supports the idea that more separable dimensions of \textit{hue-texton} lead to higher accuracy. 
Our work follows the same \changesr{ideas of quantifying integral and separable dimensions} but differs from Ware's texton selection in 
two important aspects. 
First,
the Ware study focuses on finding relationships between two 
{independent} data variables. In contrast, ours demands
participants to examine a complex scene for item discrimination when
the two variables are component parts of a vector magnitude.
Second, our texture uses the amount of black and white to show 
{luminance variations,} in contrast to the discrete shape variation in textons.
We anticipate that ours will be more suitable to continuous
quantitative values so it is easier to compare large and small to divide the regions{~\cite{wolfe2004attributes}}.
No existing work we know of has studied whether or not \changesr {one of the separable features being categorical} can facilitate global comparisons and can be scaled to \changesr{large and more complex 3D vector magnitude analysis.}

\textbf{Scene-Guidance and Feature Distance.}
\changes{In order to recognize items, viewers do not ``see'' features and instead ``bind'' these features  to objects. This binding studies how our visual systems separate object features such as shape, color, motion trajectories, sizes, and distances into the whole object~\cite{treisman1980feature}. 
}
What we ``see'' also depends on our goals
and expectations. Wolfe et al. propose the theory of ``\textit{guided
search}''~\cite{wolfe2021guided}, a first attempt to incorporate users' goals
into viewing. 
\changes{For example, if the viewer's goal is to search largest values, s/he can just check the yellow ones in Figure~\ref{fig:cases}.}
Wolfe et al.~\cite{wolfe2021guided} further suggest that
color, texture, size, and spatial frequency are among the most
effective features in attracting the user's attention.

When we combine features together, Duncan and Humphreys articulate some of the most basic principles. In general, guidance to a target will be stronger when the feature differences between the target (T) and distractor (D) are larger (TD differences), and when the feature differences amongst distractors are smaller (DD similarity)~\cite{duncan1989visual}.
\changesr{For example, Ts are 2.3 (digit) and 2 (exponent) for 230 ($2.3\times10^2$). Ds include all numbers but 2.3 and 2.}
Using the TD differences between features may explain why \textit{splitVectors} was time
consuming.
\changesr{For example, to compare 230 ($2.3\times10^2$) to 2,300 ($2.3\times10^3$),
viewers have to differentiate the two lengths of 2 (T) and 3 (T) from other distractors (Ds other than 2 or 3).
The heterogeneity of Ds or small DD distances from 3D lengths may make the use of splitVectors challenging, thus introducing temporal cost.}

\textbf{Preattentive and Attentive Feature Ranking.} \changes{Human visual processing can be faster when it is preattentive.} \changesr{Wolfe called a feature preattentive when it guides attention in search and cannot be decomposed into simpler features~\cite{wolfe2019preattentive}.}
The idea of preattentive pop-out has historically highlighted that \textit{a single object} has been considered compelling because it captures the user's attention against a background of other objects (e.g., in showing spatial  highlights~\cite{strobelt2016guidelines}).
Visual features such as orientation and color (hue, saturation, lightness) can generate pop-out effects
~\cite{healey1996high}. \changesr{This type of pop-out was also used visualizations. For example, Ropinski, Oeltze, and Preim~\cite{ropinski2011survey} summarized two groups of glyph design: \textit{``parameter mapping''} from shape and appearance (color, transparency, and texture) and ``\textit{placement}'' driven by features or data. Our study concerns appearance.
}

Recent vision science development also suggests that
the preattentive feature is not limited to single items but expanded to \textit{high-level structures}. Global statistical and structural features can be 
also preattentive~\cite{wolfe2019preattentive}.
Unlike the now outdated Treisman's 1988 preattentive processing~\cite{treisman1988feature}, where
preattentive features were considered to be 
perceived \textit{before} it is given focused attention{~\cite{treisman1988feature}},
these preattentive features \add{are} \textit{persistent during} \change{view's}{viewers'} data exploration thus can continue to provide guidance~\cite{wolfe2019preattentive, wolfe2021guided}. Viewers can use peripheral vision to compare in parallel to confidently tell apart regions relevant or irrelevant to tasks~\cite{buetti2019predicting}. 

Visual features 
also can be responsible for different attention speeds, and color (hue) and size (length and spatial frequency) are among those that guide attention{~\cite{wolfe2004attributes, ariely2001seeing}}.
Healey and Enns~\cite{healey2012attention} 
in their comprehensive review further
remark that
these visual features are not popped-out at the same speed:
\textit{hue} has higher priority than \textit{shape} and \textit{texture}~\cite{callaghan1989interference}.  
\changes{
Also, when data size
increased, some preattentive features diminished~\cite{fuchs2016systematic}~\cite{mcnabb2017survey}.
}

For visualizing quantitative data, MacKinlay~\cite{mackinlay1986automating} and Cleveland and McGill~\cite{cleveland1984graphical} leverage the 
ranking of visual features and suggest that position and size are quantitative and can be compared 
in 2D.
For example, 
\changesr{MacKinlay's A Presentation Tool (APT)~\cite{mackinlay1986automating} automatically recommends visualizations using \textit{ effectiveness} and \textit{expressive} criteria and outputs a ranked set of encoding to enumerate candidate visualizations based on data types.
}
Casner~\cite{casner1991task}
\changesr{expands} MacKinlay's APT by incorporating user tasks to guide visualization generation.
\changes{McColeman et al.~\cite{mccoleman2021rethinking}
revise the ranking of visual features based on the number of items.} 
\changesr{All these studies almost exclusively consider only single item mappings.}
Demiralp et al.~\cite{demiralp2014learning} evaluate a crowdsourcing
method to study subjective perceptual distances of 2D bivariate pairs of shape-color, shape-size, and size-color. When adopted in 3D glyph design, the authors further suggest that the most important data  
attributes should be displayed with the most salient 
visual features, to avoid situations in which secondary data values mask the information the viewer wants to see. 
\changesr{Ours also emphasizes the use of global scene features to optimize viewing experiences.}

\textbf{Monotonicity.}
Quantitative data encoding must normally 
be monotonic, and various researchers have recommended a coloring sequence that increases monotonically in luminance{~\cite{rogowitz2001blair}}. 
In addition, the visual system mostly uses luminance variation to determine shape information{~\cite{o2010influence}}. There has been much debate about the proper design of a color sequence for displaying quantitative data, mostly in 2D{~\cite{harrower2003colorbrewer}} and in
3D shape volume variations{~\cite{zhang2016glyph}}.
Our primary requirement is 
\changesr{to use categorical colormaps} that users be able to read large or
small exponents at a glance. We used four color steps 
in the first study and up to seven steps in the second study from ColorBrewer {~\cite{harrower2003colorbrewer}}
for showing areas of large and small exponents that are mapped to a hue-varying sequence. 
We claim not that {these color sequences} are optimal, only that they are reasonable solutions to the design problem. 



\section{\changesr{Experiment I: Effect of Separable Pairs on Local Discrimination and Comparison}}

The goal  in this  first experiment is to quantify the  benefits of separable pairs with preattentive features for visual  processing of a few items.  This section discusses the experiment, the design knowledge we can gain  from it, and  the factors  that  influence our design.

\subsection{Methods}
\subsubsection{Bivariate Feature-Pairs}

We chose five bivariate feature-pairs 
to examine the comparison task efficiency of separable-integral pairs.

\Lengthylengthx (\textbf{\textit{integral}}) (Figure~\ref{fig:teaser}a). 
Lengths encoded digits and exponents shown as the \change{diagonal}{height} and \change{height}{radius} of cylinder glyphs.

\Lengthycolorlengthx (\textbf{\textit{redundant and separable}}) (Figure~\ref{fig:teaser}b). This pair compared to \lengthylengthx added
a redundant color (luminance and hue variations) dimension to the exponent and  the four sequential colors were chosen from
Colorbrewer~\cite{harrower2003colorbrewer} (Appendix A shows the sequences.) 

\Lengthycolor (\textbf{\textit{separable}}) (Figure~\ref{fig:teaser}c). This pair 
mapped exponents to color. Pilot testing showed that the least 
\changesr{incorrect
exponent levels} were selected among these five feature-pairs.

\Lengthytexture (\textbf{\textit{separable}}) (Figure~\ref{fig:teaser}d). Texture represented exponents. The percentage
of black color (Bertin{~\cite{bertin1967semiology}}) was used to represent the exponential
terms $0$ ($0\%$), $1$ ($30\%$), $2$ ($60\%$) and $3$ ($90\%$), wrapped around the cylinders in  five segments to make them visible from any viewpoint.

\Lengthylengthy (\textbf{\textit{splitVectors}~\cite{henan2017}}, \textbf{\textit{separable})} (Figure~\ref{fig:teaser}e). This glyph used \change{splitVectors}{\textit{splitVectors}}~\cite{henan2017} as the baseline and mapped both digit and exponent to lengths. The glyphs were semitransparent so that the inner cylinders showing the digit terms were legible.

\textit{Feather-like fishbone legends} were added at each location when the visual variable
\textit{length} was used. The \textit{tick-mark band} was depicted as subtle light-gray lines around each cylinder. Distances between neighboring lines show a unit length legible at certain distance (Figure~\ref{fig:teaser}, rows $1$ and $2$).

\subsubsection{Hypotheses}

Given the analysis below and recommendations in the literature, we arrived at the following working hypotheses:

\begin{itemize}

\item
\textit{Exp I. H1. (Overall). The \lengthycolor feature-pair can lead to the most accurate answers.}


\item 
\textit{Exp I. H2. (Integral-separable).
Among the three separable dimensions, \lengthycolor may lead to the greatest
speed and accuracy and \lengthytexture will be more effective than \lengthylengthy (\textit{splitVectors}).}


\item \textit{Exp I. H3. (Redundancy on time).
The redundant pair \lengthycolorlengthx will reduce task completion time compared to \textit{splitVectors}}. 

\end{itemize}

Several reasons led to H1 and H2. 
\changesr{They are related to the two conditions of glyph design we evaluate.} Color and length were separable dimensions, so comparing length to color is simple \changesr{(condition 1)}.
And color was preattentive and could be detected quickly  \changesr{(condition 2)}. Compared to the redundant 
\lengthycolorlengthx, \lengthycolor reduced crowding since the {feature-pairs} were generally smaller than those in \lengthycolorlengthx. Also, distinguishing two lengths in \textit{splitVectors} might be less efficient than \lengthytexture. 
H3 could be supported because redundancy increased information processing capacity~\cite{ware2020information}. 
Redundancy contributes to efficiency by increasing the feature distances between exponents. 
\changes{We did not expect accuracy gain from redundancy because \change{splitVectors}{\textit{splitVectors}} achieved the same level of accuracy as reading texts in Zhao et al.~\cite{henan2017}. 
It may not be useful to decode quantitative data in this experiment at least for showing a few items.
}


\subsubsection{Tasks}

\begin{figure}[!bt]
	\begin{subfigure}[t]{\columnwidth}
 \centering 
 \includegraphics[width=\columnwidth]{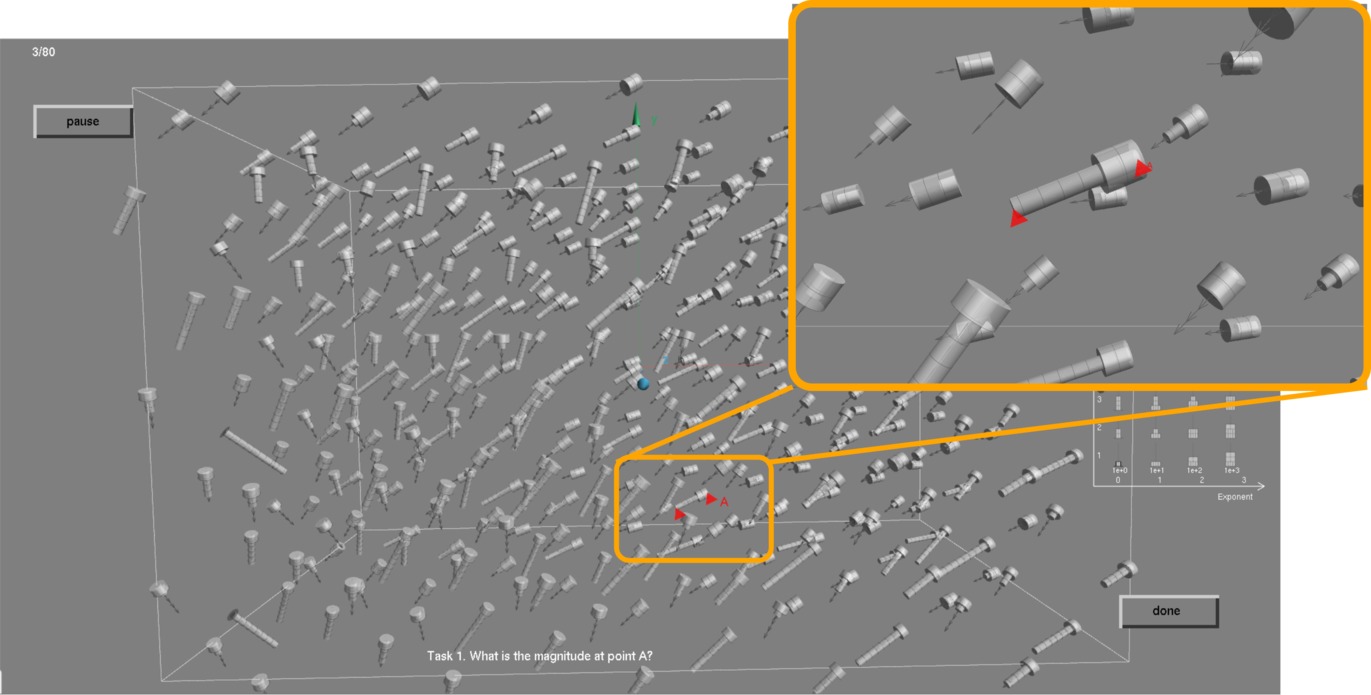}
 \caption{\remove{Exp1 }MAG task: What is the magnitude of the vector at point A? (answer: $636.30$)}
 \label{fig:task1}
    \end{subfigure}

	\begin{subfigure}[t]{\columnwidth}
 \centering 
 \includegraphics[width=\columnwidth]{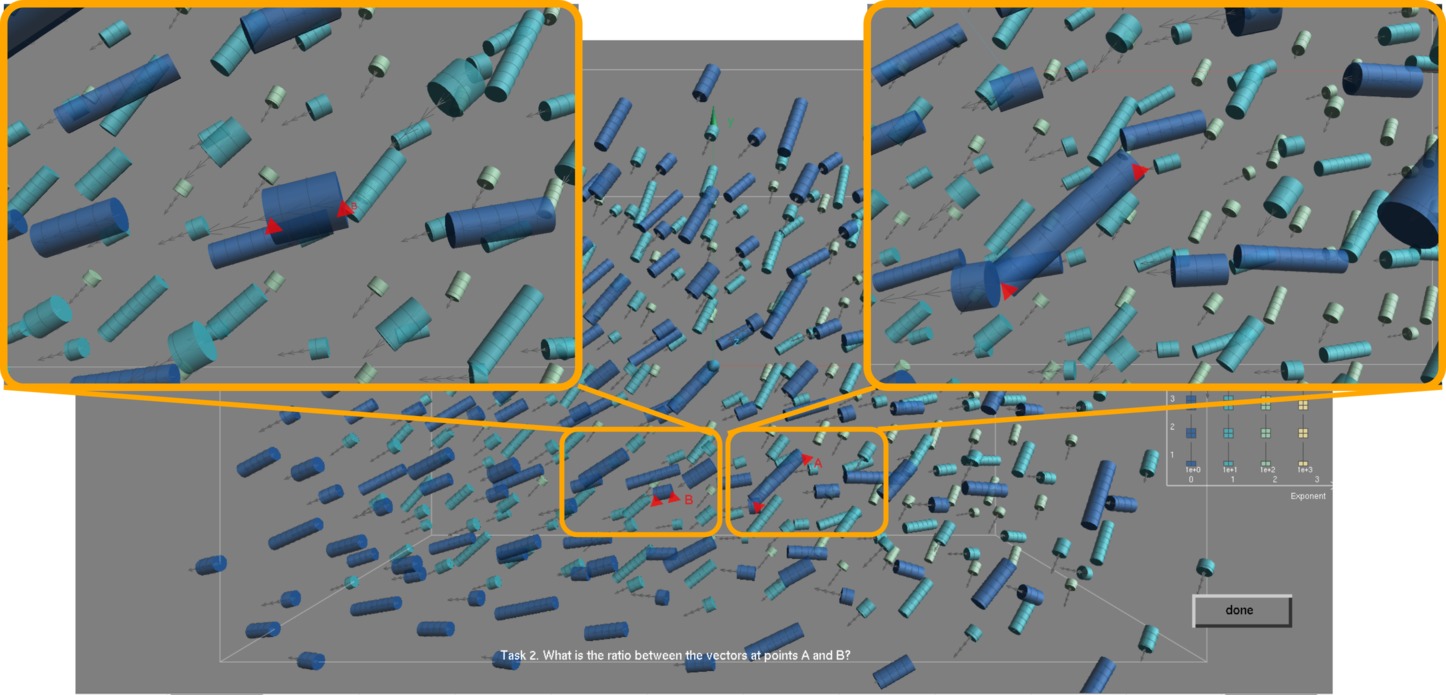}
 \caption{\remove{Exp1 }RATIO task: What is the ratio of the magnitude between the vectors at points A and B? (answer: $3.60$)}
 \label{fig:task2}
\end{subfigure}

	\begin{subfigure}[t]{\columnwidth}
 \centering 
 \includegraphics[width=\columnwidth]{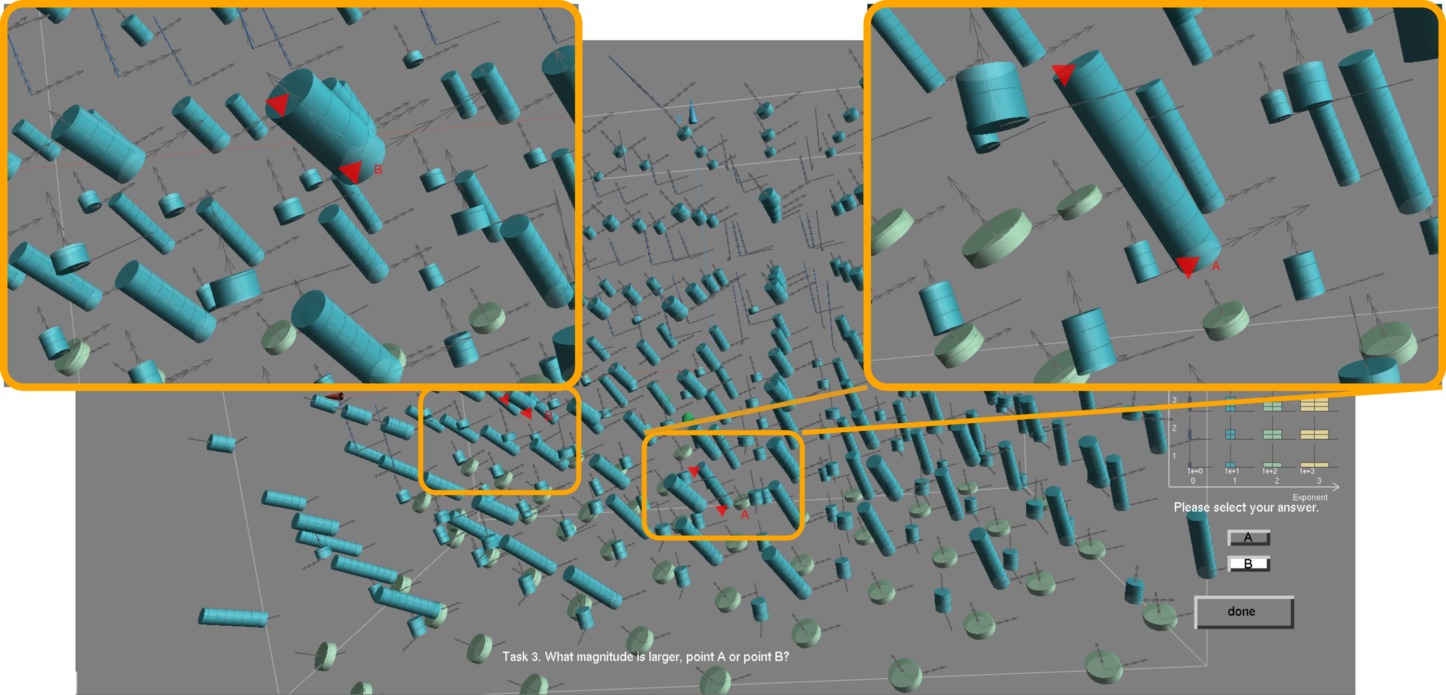}
 \caption{\remove{Exp1 }COMP task: Which magnitude is larger,  point A or point B? (answer: A on the right.)}
 \label{fig:task3}
\end{subfigure}

\caption{Experiment \change{1}{I}: Local discrimination and comparison tasks. 
\changesr{These two red equilateral triangles are rendered on the screen coordinate and are thus always visible.} 
}
\label{fig:expTasks}
\end{figure}

Participants performed the following three task types 
as in Zhao et al.{~\cite{henan2017}} so that results were comparable.
They had unlimited time to perform these three tasks.

\textbf{\change{Exp1}{Exp I}. Task 1 (MAG):  magnitude reading (Figure~\ref{fig:task1})}. \textit{What is the magnitude  at  point  A?} 
One  vector  was  marked  by  a  red  triangle labeled ``A'', and participants should report the magnitude of that vector. This task required precise
numerical input. 

\textbf{\change{Exp1}{Exp I}. Task 2 (RATIO):   ratio  estimation (Figure~\ref{fig:task2})}. \textit{What is the ratio of magnitudes of points A and B?}
Two vectors are marked with two red triangles labeled ``A'' and ``B'', and participants should estimate  the  ratio  of  magnitudes  of  these  two
vectors. 
The ratio judgment is the most challenging quantitative task~\cite{mackinlay1986automating}. Participants could either compare the glyph shapes or decipher each vector magnitude and compute the ratio mentally.

\textbf{\change{Exp1}{Exp I}. Task 3 (COMP):  comparison (Figure~\ref{fig:task3})}. \textit{Which magnitude is larger,  point A or B?} 
Two vectors are marked with red triangles and labeled ``A'' and ``B''. Participants select their answer by directly clicking the ``A'' or ``B'' answer buttons. This task was a simple comparison between two values and offered a binary choice of large or small.

\subsubsection{Data Selection}
\label{sec:dataselection}

Because we were also interested in comparing our results to those in Zhao et al.{~\cite{henan2017}}, we 
replicated their data selection method
by randomly sampling some quantum physics simulation results and produce samples within 3D boxes of size $5\times3\times3$.
There were $445$ to $455$ sampling locations in each selected data region.

We selected the data satisfying the same following conditions: (1) the answers must be at locations where some context information was available, i.e., not too close to the boundary of the testing data; (2) no data sample was repeated to the same participant; 
(3) since data must include a broad measurement, we selected the task-relevant data from 
each exponential term of $0$ to $3$.

\subsubsection{Empirical Study Design}


\textbf{Design and Order of Trials.} We used a within-subject design with one independent variable of bivariate quantitative {feature-pair} (five types).
Dependent variables were error
and task completion time. We also collected participants' confidence levels.
Table~\ref{tab:experimentdesign} showed that participants were assigned into five blocks in
a Latin-square order, and within one block the order of the five {feature-pair} types is the same.
Participants performed tasks with randomly selected datasets.
Each participant performed 
$60$ {trials} ($3$ tasks $\times$ $4$ random data $\times$ $5$ {feature-pairs}). 
These four random data were from four exponent ranges.

\textbf{Participants.}
We diversified the participant pool as much as possible, since all tasks could be carried out by those with only some science background. 
Twenty participants ($15$ male and $5$ female, 
mean age $=$ $23.3$, and standard deviation $=$ $4.02$) participated in the
study, with ten in computer science, three in engineering,
two in chemistry, one in physics, one in linguistics, one
in business administration, one double-major in computer
science and math, and one double-major in biology and
psychology. The five females were placed in each of the five
blocks (Table~\ref{tab:experimentdesign}). 
On average, participants spent about $40$ minutes on the 
tasks.

\begin{table}[!t]
\caption{
Experiment I design: $20$ participants are assigned to one of the five blocks and use all five bivariate pairs. 
Here, \lyly: \lengthylengthy (\textit{splitVectors}), \lylx: \lengthylengthx, \lc: \lengthycolor, \lt: \lengthytexture,
and \lcl: \lengthycolorlengthx.
}
\label{tab:experimentdesign}
\scriptsize
\begin{center}
\begin{tabular}{ l l l }
   \toprule
 \centering Block & Participant & Feature-pair   \\
 \midrule
 	1 & P1, P6, P11, P16 & \textit{splitVectors}, \lylx, \lc, \lt, \lcl\\
    2 & P2, P7, P12, P17 & \lylx, \lc, \lt, \lcl, \textit{splitVectors}\\
    3 & P3, P8, P13, P18 & \lc, \lt, \lcl, \textit{splitVectors}, \lylx\\
    4 & P4, P9, P14, P19 & \lt, \lcl, \textit{splitVectors}, \lylx, \lc\\
    5 & P5, P10, P15, P20 & \lcl, \textit{splitVectors}, \lylx, \lc, \lt\\ 
  \bottomrule
\end{tabular}
\end{center}
\end{table}

\textbf{Procedure.}
Participants  were  greeted  and  completed  an  Institutional  Review Board (IRB) consent form (which described the procedure, risks and benefits of the study) and the demographic survey. All participants had  normal or corrected-to-normal vision and passed the Ishihara color-blindness test. 
We showed {feature-pair} examples and trained the participants with one trial for {every feature-pair} per task.  They were told to be as accurate and as quickly 
as possible, and that accuracy was more important than time. They could ask questions during the training but were told they could not do so during the formal study. Participants practiced until they fully understood the {feature-pairs} and tasks. After the formal study, participants filled in a post-questionnaire asking how these feature pairs supported their tasks and were interviewed for their comments. \changesr{Pilot studies were conducted to examine the procedures.}

\textbf{\changes{Environment.}}
Participants  sat  at  a $27\,''$ BenQ  GTG  XL  $2720$Z, gamma-corrected display with resolution $1920$ $\times$ $1080$ to ensure the colors were displayed properly. The distance  between  the  participants  and  the  display  was  about $50$cm. The minimum visual angle of task-associated glyphs was $0.2^{\circ}$ in the default view where all data points were visible and the scene filled the screen. 

\textbf{\changes{Interaction.}}
Participants could rotate the data and zoom in and out. Lighting placement and intensity were chosen to produce visualization with contrast and lighting properties appropriate for human assumptions and the spatial data. The screen background color was neutral stimulus-free gray background to minimize the discriminability and appearance of colors~\cite{ware2020information}. Using black or white background colors makes the black and white texture stimuli disappear thus bias the results \changes{(See Appendix B for examples)}.

\subsection{Experiment I: Results and Discussion}

\subsubsection{Analysis Approaches} 

We collected $400$ data points for each task. 
In preparing the accuracy and task completion time for analysis,
\changesr{we differentiated two error metrics related to the perceptual accuracy of the bivariate pairs:}
\begin{itemize}
    \item
    \changesr{
{Correspondence error (C-Error)}: 
A trial is considered to have an answer of C-Error \changes{if} response's \textit{exponent} value does not match the correct one.
Having a C-Error would mean that participants have trouble differentiating the exponent levels within a glyph. }

\item
\changesr{
{Relative error (R-Error)}: 
This R-Error follows Zhao et al.{~\cite{henan2017}} to study how sensitive a method is to error uncertainty based on fractional uncertainty, calculated as \textit{R-Error = $\lvert$ correct answer - participant answer $\rvert$ / (correct answer)}. This measure was used for MAG and RATIO tasks. The benefit of this metric was that it took into account the value of the quantity being compared and thus provided an accurate view of the overall errors.
}
\end{itemize}

\changesr{
In subsequent analysis, 
we  separated these two error measurements since
Combining these two errors in the analysis would also be problematic. The C-Errors are at least one order of magnitude larger or smaller than the ground truth. 
We also did not remove participants' data with C-Errors, since the source of errors was caused by glyph design methods independent of trials. 
}


A post-hoc analysis using Tukey's Studentized Range test (HSD) was performed when we observed a significant main effect on R-Errors. 
When the dependent variable was binary (i.e., answer correct or wrong), we used a logistic regression and reported the \textit{p} value from the Wald $\chi^2$ test.
When the $p$ value was less than 0.05, variable levels with $95\%$ confidence interval of odds ratios not overlapping were considered significantly different.
All error bars represent $95\%$ confidence intervals.
We also evaluated effect sizes 
{using \textit{eta-square}, labeled ``small'' $(0.01-0.06)$, ``medium'' $[0.06-0.14)$, and ``large'' $(\geq 0.14)$ effects following Cohen~\cite{cohen1988statistical}.
}

\begin{table}[!tp]
	\caption{Summary statistics by tasks. 
    The significant main effects and the high
	effect size (ES) are in \textbf{bold} (none in these observations) and the medium effect size is in \textit{italic}. Effect size is 
eta-square 
labeled ``small'' $(0.01-0.06)$, ``medium'' \change{$[0.06-0.14)$}{$[0.06-0.14)$}, and ``large'' \change{$\geq 0.14$}{$(\geq 0.14)$}
effects following Cohen~\cite{cohen1988statistical}.	Post-hoc Tukey grouping results are reported for significant main effects, where $>$ means statistically significantly better and enclosing parentheses mean they belong to the same Tukey group.
}
	\label{tab:new_glm}
	\scriptsize
	\begin{center}
		\begin{tabular}{l l l l}
		\toprule
			\centering Task & Variables & Significance & ES\\
			\midrule
	MAG &  time  & \textbf{F$_{(4,\,384)}$ = 6.8, \textit{p} $<$ 0.0001} & \textit{0.07} \\ 
	    & & \textbf{(\lc, \lt, \lcl, \textit{splitVectors})} $>$ \textbf{\lylx} &\\
			& relative error & F$_{(4,\,384)}$ = 0.9, \textit{p} = 0.46 & 0.01 \\ 
			\midrule
	RATIO  &  time  &  \textbf{F$_{(4,\,395)}$ = 6.2,  \textit{p} $<$ 0.0001} & \textit{0.06} \\ 
	& & Three groups: \textbf{A: \lc, \textit{splitVectors}, \lt} &\\
	& & \hspace{1.53cm} \textbf{B: \textit{splitVectors}, \lt, \lcl} & \\
	& & \hspace{1.53cm} \textbf{C: \lt, \lcl, \lylx}& \\
	& relative error & F$_{(4,\,395)}$ = 0.8, \textit{p} =  0.50 & 0.01 \\ 
	\midrule
	COMP  &  time  &  \textbf{F$_{(4,\,395)}$ = 10.4, \textit{p} $<$ 0.0001} & \textit{0.09}\\ 
	& & Three groups: \textbf{A: \lcl, \lc, \lt}& \\
	& & \hspace{1.53cm} \textbf{B: \lc, \textit{splitVectors}} & \\
	& & \hspace{1.53cm} \textbf{C: \textit{splitVectors}, \lylx} & \\
	& accuracy &  ${\chi}^2$ = 0.4, \textit{p} = 0.98 & 0.03\\
	
			\bottomrule
		\end{tabular}
	\end{center}
\end{table}

\begin{figure*}[!thp]
\centering
	\begin{subfigure}{0.33\textwidth}
    \includegraphics[width=\textwidth]{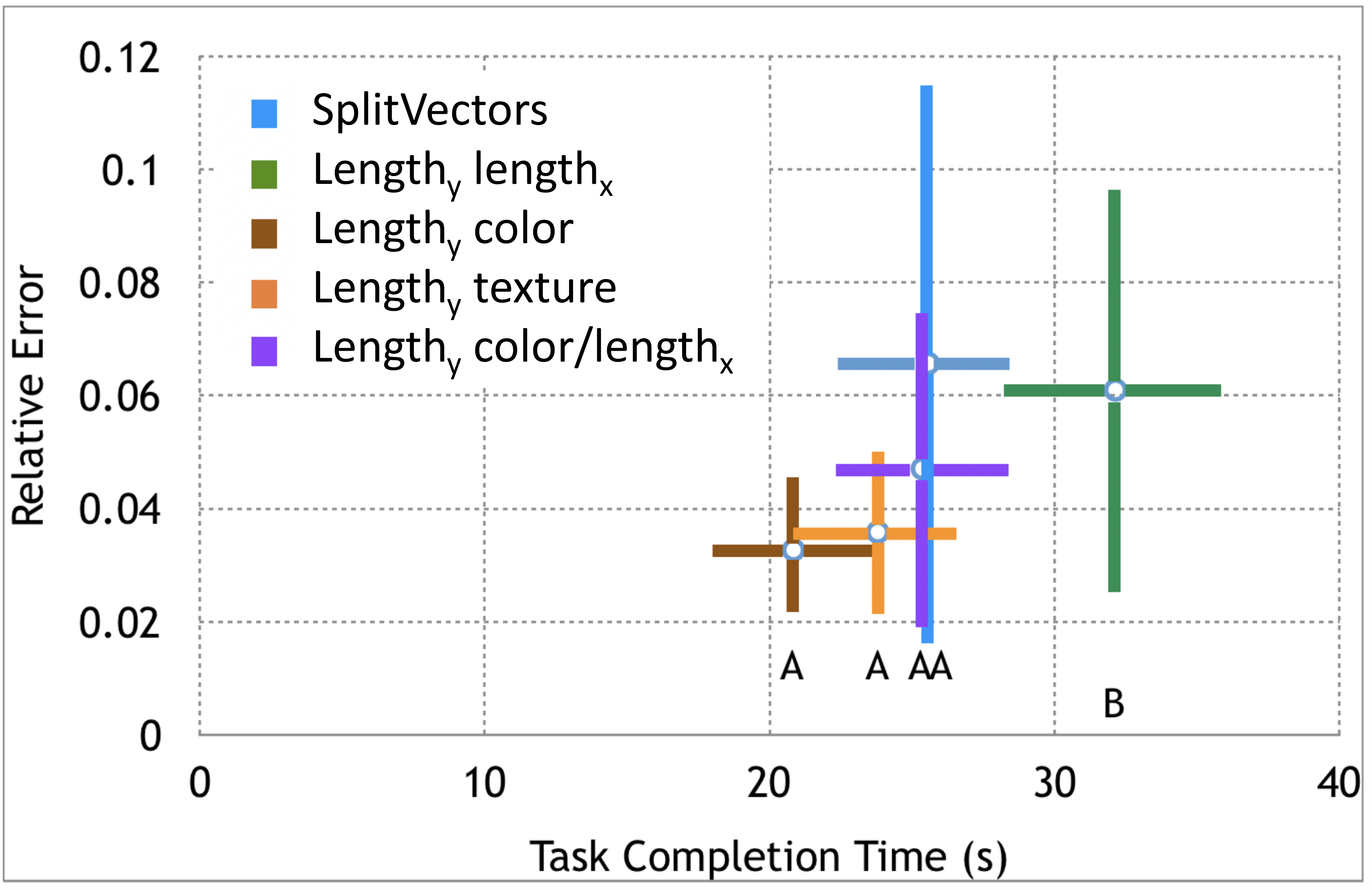}
		\caption{Task 1 (MAG)}
		\label{fig:task1timeError}
	\end{subfigure}
\begin{subfigure}{0.33\textwidth}
		\centering
	\includegraphics[width=\textwidth]{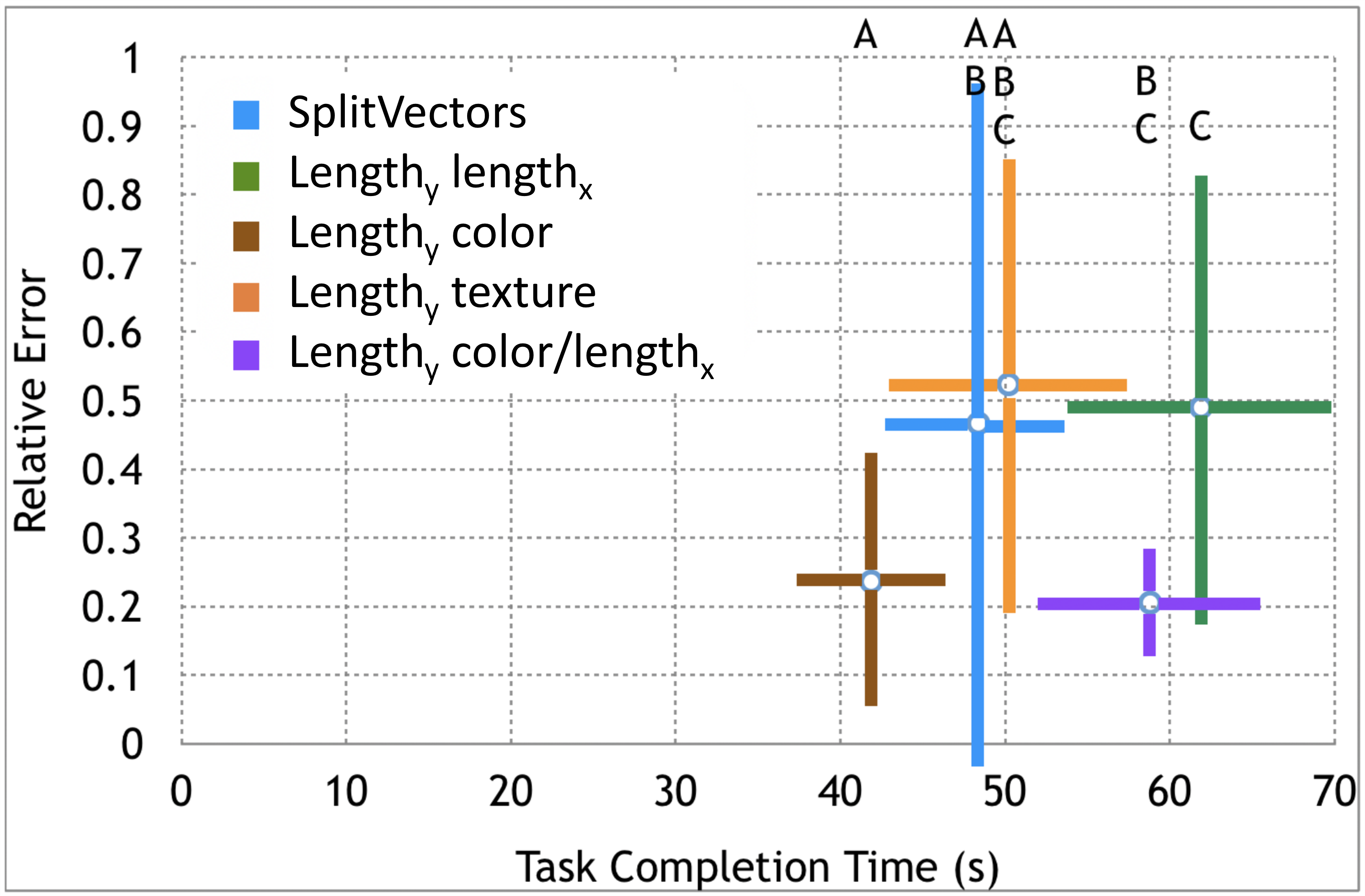}
		\caption{Task 2 (RATIO)}
		\label{fig:task2timeError}
	\end{subfigure}
	\begin{subfigure}{0.33\textwidth}
	\includegraphics[width=\textwidth]{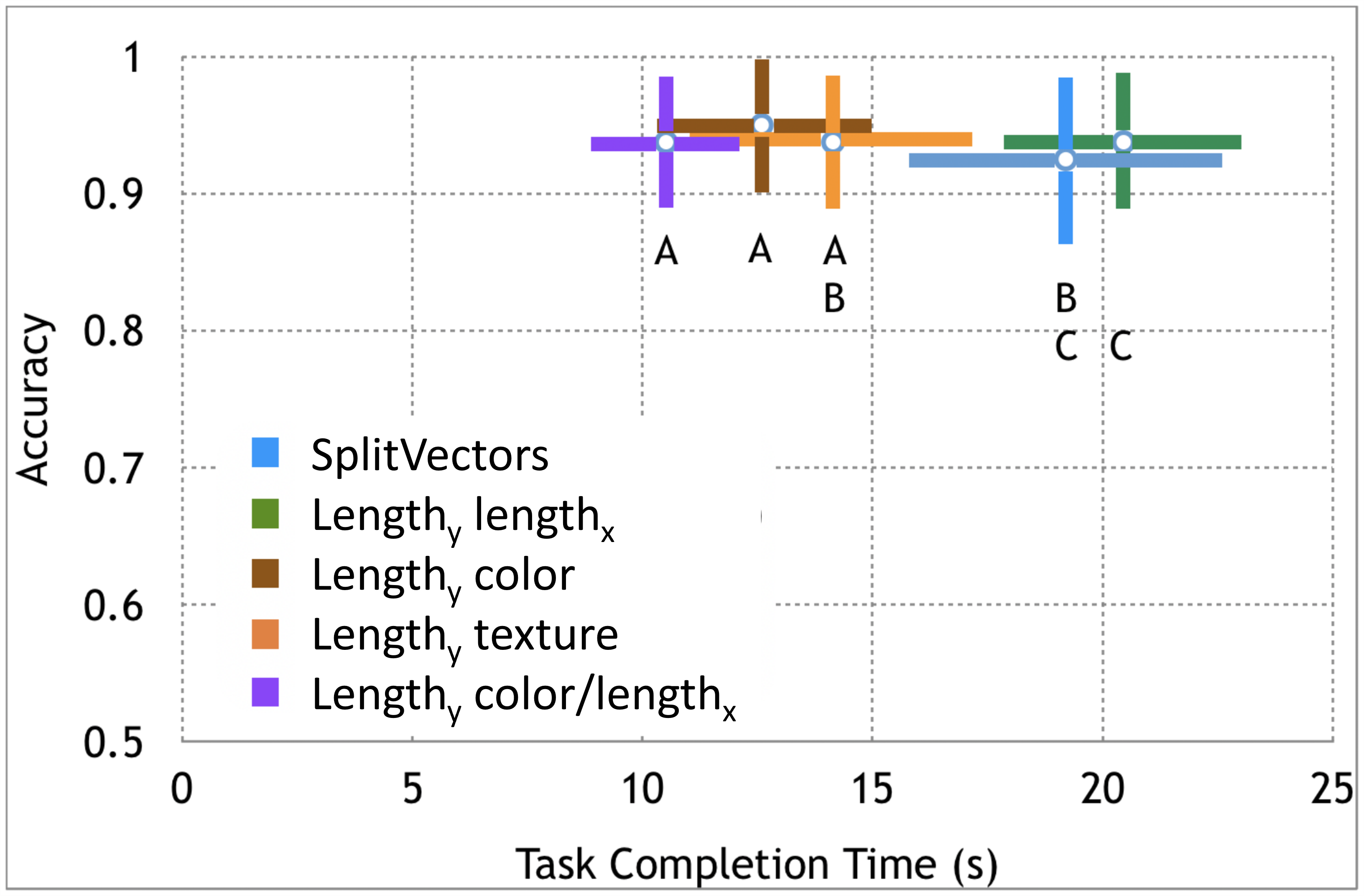}
		\caption{Task 3 (COMP)}
		\label{fig:task3timeError}
	\end{subfigure}

\caption{Experiment I task completion time and relative error or accuracy by tasks. 
The horizontal axis represents the mean task completion time while the vertical axis showing the accuracy or relative error. Same letters represent the same post-hoc analysis group. Colors label the feature-pair types. All error bars represent $95\%$ confidence interval.
}

\label{fig:timeError}
\end{figure*}



\subsubsection{Overview of Study Results}

Figure~\ref{fig:exp1cerror} show all C-Error occurrences. 
Table{~\ref{tab:new_glm}} and Figure{~\ref{fig:timeError}} show the $F$ and $p$ values computed with SAS one-way measures of variance for task completion time and relative error. 
Our results clearly demonstrated the benefits  
in terms of task completion time
of separable
dimensions for comparison. 
We observed a significant main effect of feature-pair type on task completion time for all three tasks MAG, RATIO, and COMP, and the effect sizes were in the medium range. 
\Lengthycolor was the most efficient approach.
For 
{COMP}, \lengthycolor, \lengthytexture and \lengthycolorlengthx were 
most efficient for simple two-point comparisons (Figure{~\ref{fig:task3timeError}}).

\subsubsection{Separable Dimension Coupled with Categorical Features had the Least Correspondence Errors.}
\label{sec:cerror}

\begin{figure}[!tb]
\centering
    \includegraphics[width=\columnwidth]{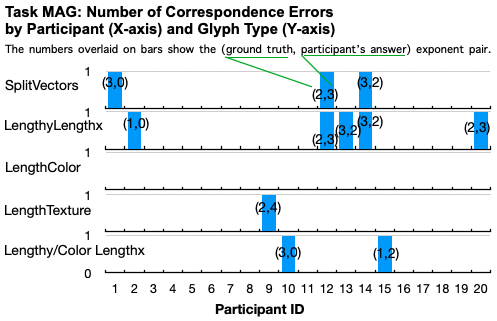}
	\caption{\changes{Experiment I (Task MAG): All instances of correspondence errors by participant.
	The most separable \lengthycolor glyph had no instances of correspondence error whilst the \lengthylengthx had the most.
The redundant color dimensions helped removed some correspondence
errors (Two instances of \lengthycolorlengthx vs. five instances of \lengthylengthx).}
	}
	\label{fig:exp1cerror}
\end{figure}

\changesr{We only observed C-Errors in MAG, but not in the RATIO and COMP tasks. The total count was relatively small ($11$ instances of 400 data points). They came from $9$ participants (error mean = $1.22$  and $95\%$ confidence intervals (CI)=$[0.96, 1.48]$). Figure~\ref{fig:exp1cerror} shows all instances of these errors by participant and by encoding methods.  
It appeared that the degree of separability of integral-separable dimensions influenced the errors: the most integral dimension \lengthylengthx had the highest number ($5$ instances) of C-Errors  and the most separable
\lengthycolor had none. 
}

\subsubsection{Separable Dimensions Are Better Than Integral Dimensions for Local Comparisons. But Categorical Feature was not a Statistically Significant Effect.}

\textit{Our first two hypotheses H1 and H2 are supported.}  In the MAG task, the integral \lengthylengthx was the least efficient and all other separable-pairs were in a separate group, the most efficient one (Figure~\ref{fig:task1timeError}).  In 
RATIO, 
\lengthycolor, \lengthytexture, and \textit{splitVectors} were the most efficient group (Figure~\ref{fig:task2timeError}); in 
COMP,
the redundant \lengthycolorlengthx,  \lengthycolor, and \lengthytexture were in the most efficient group (Figure~\ref{fig:task3timeError}). 
\textit{SplitVectors} was not as bad as we originally thought in perceiving correct exponents.
\textit{SplitVectors} belonged to the same efficient post-hoc group as \lengthycolor and \lengthytexture for 
RATIO 
and these three were also 
most efficient for MAG.

\subsubsection{Separable Pairs of \Lengthycolor And \Lengthycolorlengthx Achieved Comparable Efficiency To Direct Linear Glyph}
\changesr{One aspect  
for motivating this experiment was to quantify the benefits of separable pairs~\cite{borgo2013glyph, ware2020information}:
whether the separable pairs supported {COMP} and how the separable pairs compared in efficiency to the direct mapping (\change{Fig.}{Figure}~\ref{fig:cases}(f)).} 
Since our study had the same numbers of sample data as Zhao et al.~\cite{henan2017},
we then performed a one-way $t$-test 
to compare against the direct linear encoding in Zhao et al.~\cite{henan2017}. 
Our results indicated that results for COMP (judging large or small) from separable variables was no more time-consuming than direct linear glyphs, and our post-hoc analysis 
showed that \lengthycolor, \lengthycolorlengthx, and $linear$ were in the same post-hoc group.
We also observed that \textit{splitVectors} dropped to the least efficient post-hoc group (Figure~\ref{fig:task3timeError}). 
This result replicated the former study results in Zhao et al.~\cite{henan2017} 
by showing that \textit{splitVectors} impaired comparison efficiency.

\subsubsection{\changes{Redundant Feature-Pairs 
Were Efficient}}

We also confirmed hypothesis H3. We were surprised by the large performance gain with the redundant 
encoding \lengthycolorlengthx of mapping $color$ and $length$ to the exponents in \textit{splitVectors}. With the redundant encoding, the task completion time was significantly shorter than \lengthylengthx for MAG and COMP tasks. 
While Ware~\cite{ware2020information} confirmed that the efficiency might not be improved by using separable dimensions, in our case, where color and size (separable) represent the same quantitative value, we suggested that the redundancy worked because participants could use either length or color in different task conditions. We could also consider that \lengthycolorlengthx is a redundant encoding of \lengthycolor, and those two feature-pairs had similar efficiency and accuracy for all local tasks. 

\subsection{Summary}

\changesr{The separable-pair condition is necessary for effective glyph design because all separable pairs were more efficient than the integral ones. The pre-attentive condition enabled by categorical encoding among the separable pairs may be not since not all conditions were statistically different performance-wise.}
All tasks (MAG, RATIO, and COMP) lacked of significant main
effect on relative errors (in MAG or RATIO) or accuracy (in COMP).
Note that none of these three tasks required initial visual search, and target answers were labeled. Wolfe called this type of task-driven with known target guided tasks{~\cite{wolfe2021guided}}. \Lengthycolor was the most accurate in all tasks. 
 

\changesr{We also did not see the needs for the second condition for perceptually accurate glyphs in this experiment.
We did not observe differences among categorical dimensions color, texture, and length. 
We suspect that the reason 
for this lack of significance could well be their similarities 
in mentally computing load.
The load was relatively small when comparing two values.
We suspected that when search-space set-size increases, and when tasks are more complex involving all items, 
participants will need preattentive global scene features to guide their search. 
We subsequently ran the second experiment to increase the set size in tasks to the entire scene to study the benefits of categorical features to show quantitative exponent values to benefit global search. 
}

\section{\changes{\change{Empirical Study}{Experiment} II: Scalability of Global Scene Features}}

The goal in this second experiment is 
to quantify the benefits of separable feature-pairs when they introduce \changesr{categorical features of} scene guidance for \textit{global} tasks in search spaces, as large as the entire dataset of several hundreds items. \changesr{In other word}, we \changesr{measure} scene feature scalability of global tasks.

\subsection{Overview}
\label{sec:twoconsiderations}
\changesr{
We had three design considerations for us to carefully choose the categorical features in setting up this experiment, concerning the use of glyphs for showing complex simulation results. Intriguingly, all of these considerations support our second glyph design consideration of using a categorical variable in one of the separable pairs.
}

The first reason is that the initial \textit{at-a-glance} global statistical summary of the scene depends on categorical information~\cite{wolfe2019preattentive}.
\changesr{
One of the most important advances in vision science is to find that viewers can summarize the scene without attending to the specific items~\cite{whitney2018ensemble}. Visual dimensions facilitating this summary process become global scene features and these features are pre-attentive~\cite{wolfe2021guided}. While visualization is mainly about mapping data values to visual variables, the new theory concerns how features form the structural and content of the scene that can affect efficiency.
If the quantum spins contain one 
object  at  a  time,  then  the first condition of glyph design considering integral and separable dimensions is sufficient to explain the experience as we have shown in Experiment I. 
For complex tasks, in general, our visual system has a limited capacity. To cope with this limit, humans first visually summarize the scene to find specific regions of interests~\cite{borgo2013glyph, wolfe2021guided}. If categorical features stimulate population responses from multiple items, we should observe fewer errors and better efficiency.
For example, we have exemplified in the Introduction section for search of ``largest'' values by looking up ``yellow'' regions, without attending to every single items of ``yellow''.
}

\begin{figure}[!tp]
	\centering
	\includegraphics[width=\linewidth]{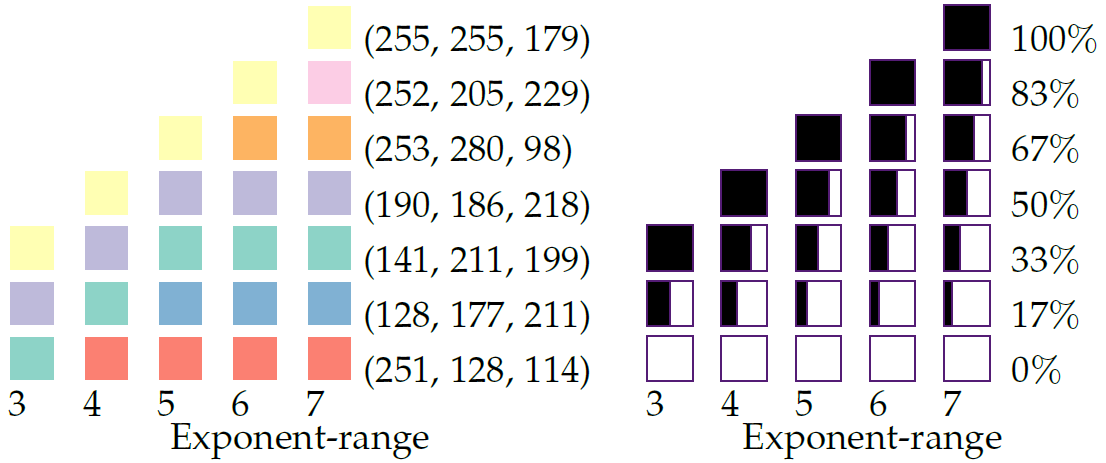}
	\caption{\changesr{Visual mapping using color and texture in Experiment II. From the top to bottom, colors and texture segments are mapped to exponent values from the largest to the smallest. The three numbers next to the 7-level colormap are the RGB values. The numbers next to the texture columns are the proportion of black-on-white for the last 7-level texture configuration.}
	}
	\label{fig:exp2VisualMapping}
\end{figure}

The second concerns \textit{scalability to feature distances}. \changesr{Here feature distance is meant to represent target-distractor similarity. It is not the absolute features (e.g., yellow) that direct our attention towards the answer; rather, what determines performance is the result of a comparison between target (yellow) and other data features (such as pink and orange) in the scene (e.g., yellow is different from other colors and the yellow regions stand out)~\cite{wolfe2021guided}.}
In other words,
one must also look at feature distractors~\cite{acevedo2007modeling, chung2016ordered, urness2003effectively}, whether or not they are heterogeneous, and that 
the efficiency  of a scene guidance will decline as a function of the degree of distractor variation~\cite{duncan1989visual, buetti2019predicting, lleras2020target}.
While generally, 
subjective reports from Experiment I 
indicate that  \lengthycolor  and \lengthytexture show the similar perceptual speed.  \changesr{Performance of texture may decline faster than  color  as  the  exponent range increases because our vision is not as sensitive to luminance-variation as to hues.
For example, at the exponent-range of 7 in \autoref{fig:exp2VisualMapping}, the differences between yellow and pink could be more differentiable than the two top-level textures of different amount of black. 
}
In this study, we expanded the  data  range from  the  single  level  in Experiment I to  five  ranges $\in[3, 7]$ to  understand feature-pair scalability to feature distances. 
The  efficiency of color in Experiment I could well  arise because the  range  (of $4$) was  not  
large  enough.

\begin{figure*}[!tb]
	\centering
	
	\begin{subfigure}[b]{\columnwidth}
		\centering
		\includegraphics[width=0.85\columnwidth]{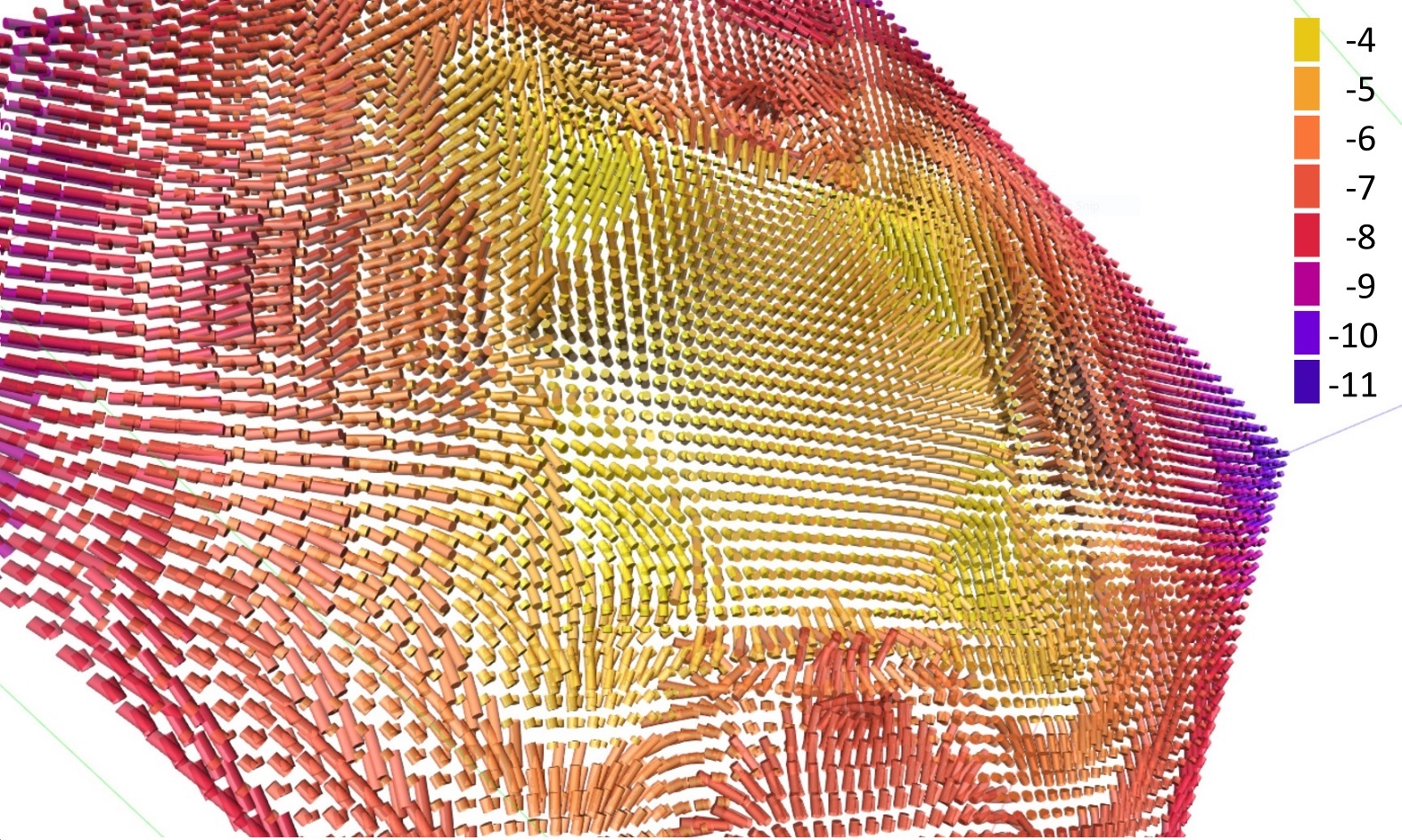}
		\caption{Continuous colormap and high-density data}
		\label{fig:continousHigh}
	\end{subfigure}
	\begin{subfigure}[b]{\columnwidth}
		\centering
		\includegraphics[width=0.85\columnwidth]{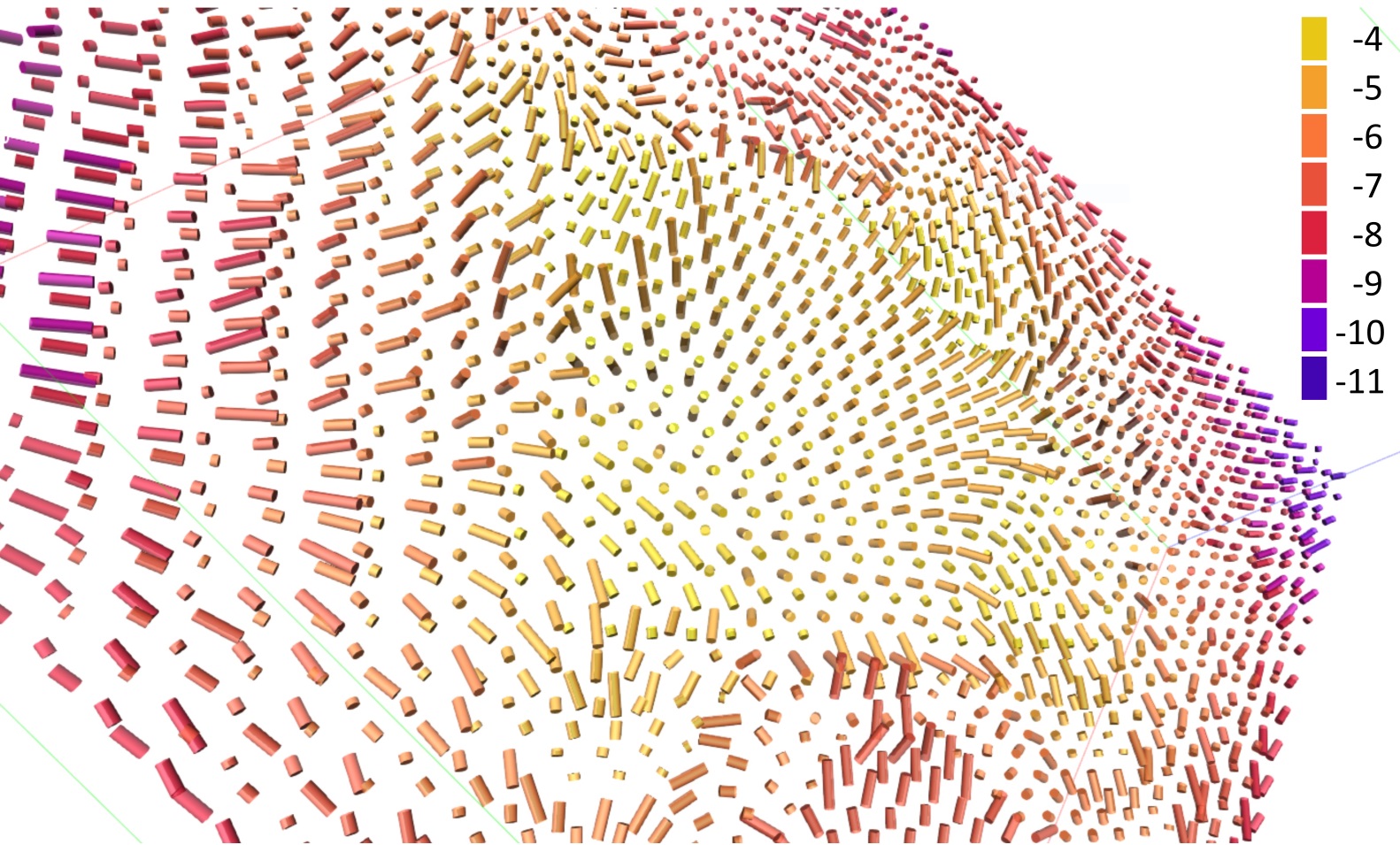}
		\caption{Continuous colormap and low-density data}
		\label{fig:continousLow}
	\end{subfigure} 
	
	\begin{subfigure}[b]{\columnwidth}
		\centering
		\includegraphics[width=0.85\columnwidth]{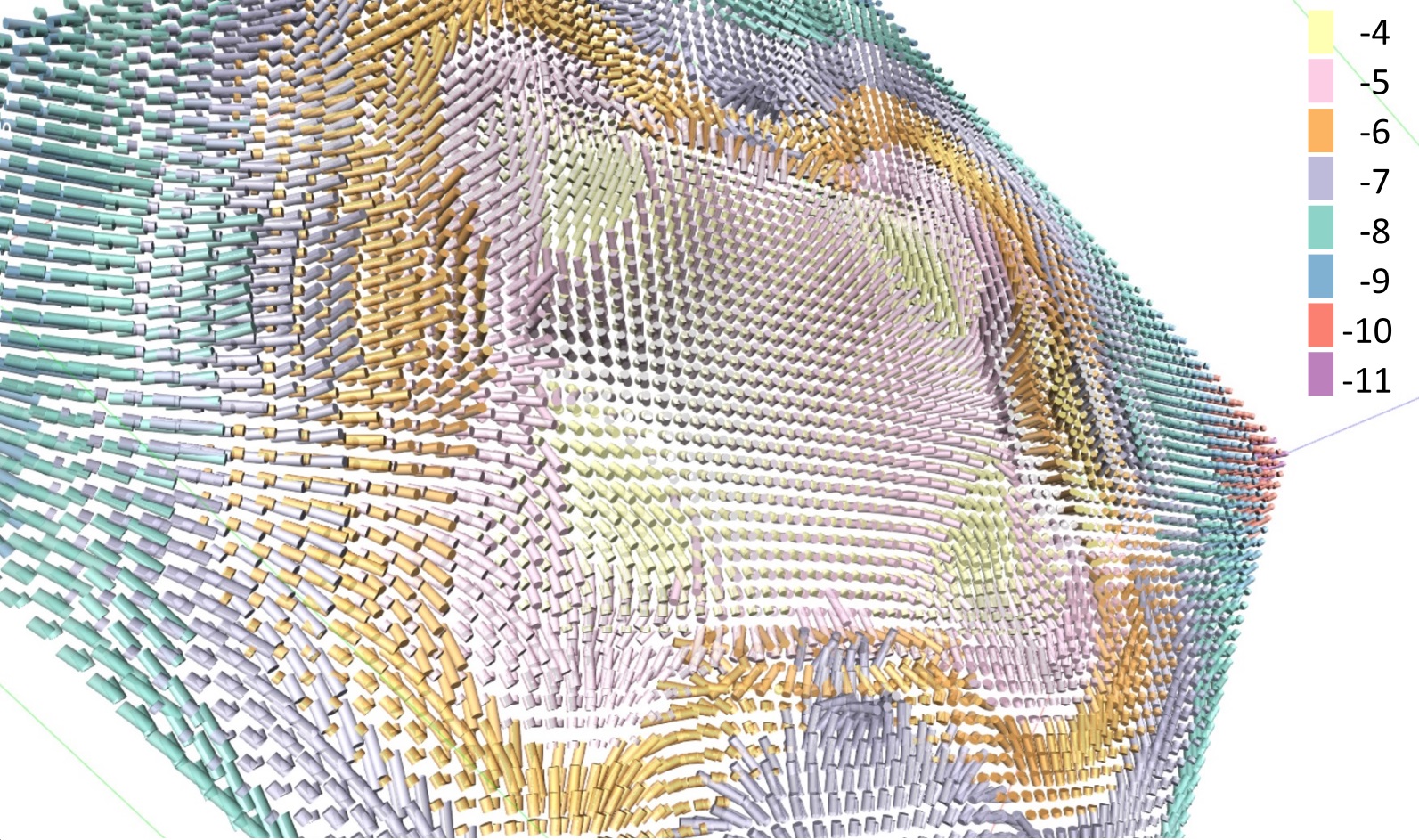}
		\caption{Categorical colormap and high-density data}
		\label{fig:catogoricalHigh}
	\end{subfigure} 
	\begin{subfigure}[b]{\columnwidth}
		\centering
		\includegraphics[width=0.85\columnwidth]{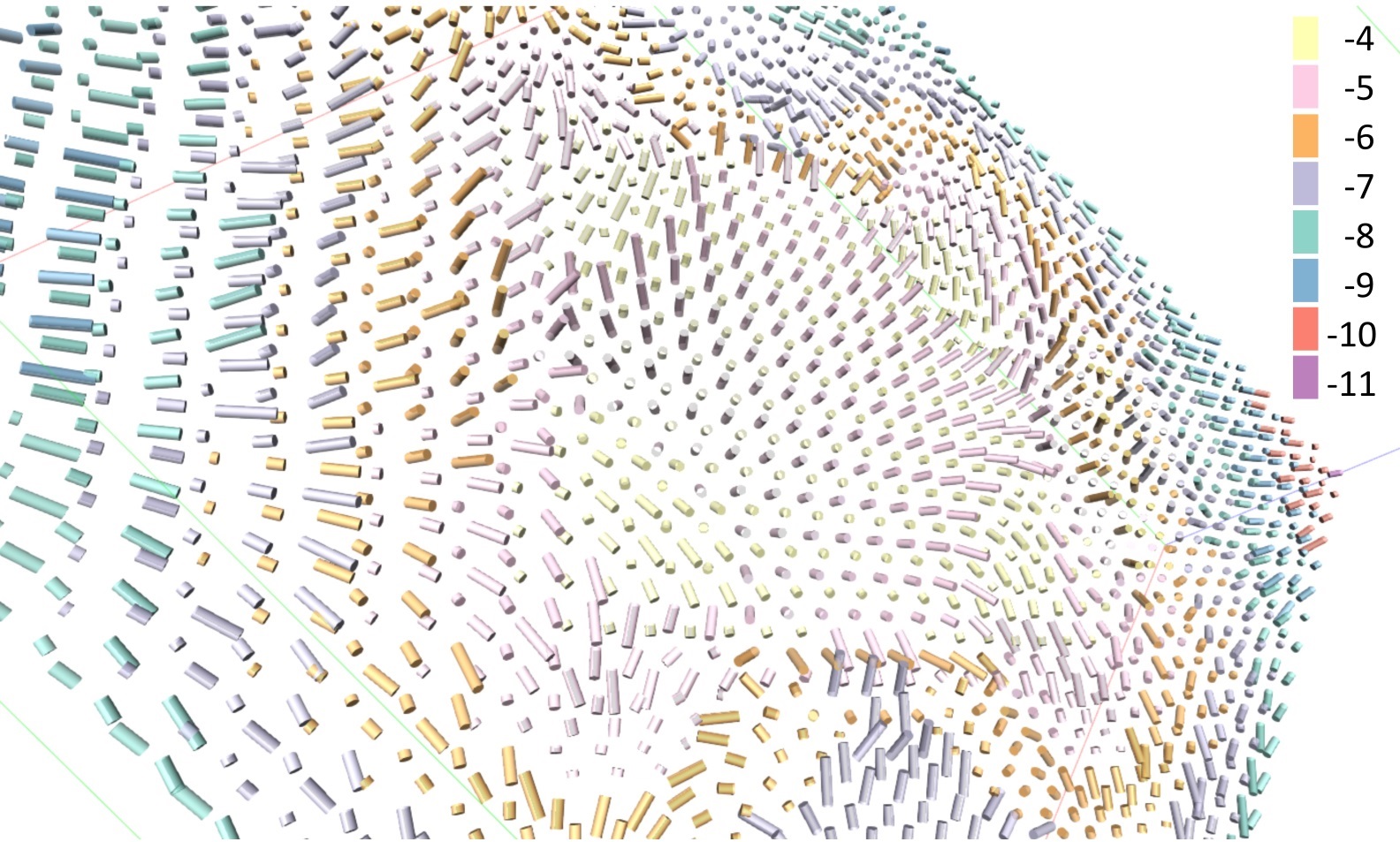}
		\caption{Categorical colormap and low-density data}
		\label{fig:catogoricalLow}
	\end{subfigure}

	\caption{\changesr{Density effects on color choices to justify the use of dense sampling and categorical colormap (c) in Experiment II. 
This example dataset shows \textit{two colormaps}: ( {segmented-continuous} (a and b)  and {categorical} (c and d) colormaps), at \textit{two different data densities. (a) and (c) show data with the raw density from the simulation results; (b) and (d) were produced by removing around $70\%$ vector glyphs.}
The boundaries between the data categories are more recognizable when the data are dense in (a) and (c) (comparing the 1st column and the 2nd column). At the same density (comparing the 1st and 2nd row), the boundaries between levels are easier to recognize when spin vectors are rendered using a categorical colormap of (c) and (d).}
We thus use the raw dense and categorical colormaps (c) in Experiment II.
}
	\label{fig:colorAndData}
\end{figure*}

The third concerns the density effects on color choices. \changesr{
Figure{~\ref{fig:colorAndData}} 
shows two densities and two colormaps (a categorical colormap from Colorbrewer~\cite{harrower2003colorbrewer} and a segmented continuous colormap by the number of exponents generated from the extended blackbody colormap).}
For a feature to actually \textit{guide} attention, 
we can see from Figure{~\ref{fig:colorAndData}}, the boundary detection with these colormaps is associated with data density.
Unless the data density was reasonably high, detecting the boundaries using continuous colormaps (Figures{~\ref{fig:continousHigh}}, {~\ref{fig:continousLow}}) is harder than the ColorBrewer colormaps (Figures{~\ref{fig:catogoricalHigh}}, {~\ref{fig:catogoricalLow}}). 

\subsection{Method}

\subsubsection{Feature-Pairs}

We used \lengthycolor, \lengthytexture, and baseline \textit{splitVectors} in Experiment II. 
These three visualizations were chosen because \lengthycolor and \lengthytexture are among the best feature-pairs from Experiment I and because color and texture are among the most separable features according to Ware{~\cite{ware2020information}}.
To introduce a 
``distractor'' experience to measure \textit{scalability to feature distances}, we vary the data range from the $4$ levels in Experiment I to $3-7$ levels in Experiment II (See mapping in \changes{Figure{~\ref{fig:powerLevelExample}}, Appendix C.\changesr{)}}

%

\subsubsection{Hypotheses}

We had the following hypotheses:

\begin{itemize}
\item

\textit{Exp II.H1 (Accuracy). \changesr{More categorical feature in the separable pairs will be more effective.} We thus  anticipate a rank order of effectiveness from high to low: \lengthycolor, \lengthytexture, and \change{$splitVectors$}{\textit{splitVectors}}.}

\item

\textit{Exp II.H2 (Correspondence Errors).}
\textit{\changesr{More categorical feature of color in the separable pairs will reduce
C-Errors,} when participants will choose the correct exponent level.}

\item
\textit{Exp II.H3 (User behavior).}
\textit{More \changesr{categorical dimension} in the separable feature-pairs will lead to optimal users' behaviors: i.e.,~participants can quickly locate  task-related regions for tasks that demand looking among many vectors due to global scene features.}

\end{itemize}



\subsubsection{Tasks}

\begin{figure}[tb]
\centering
          \begin{subfigure}[t]{0.45\textwidth}
        		\centering
        		\begin{tikzpicture}
        		\node (img1){\includegraphics[width=\textwidth]{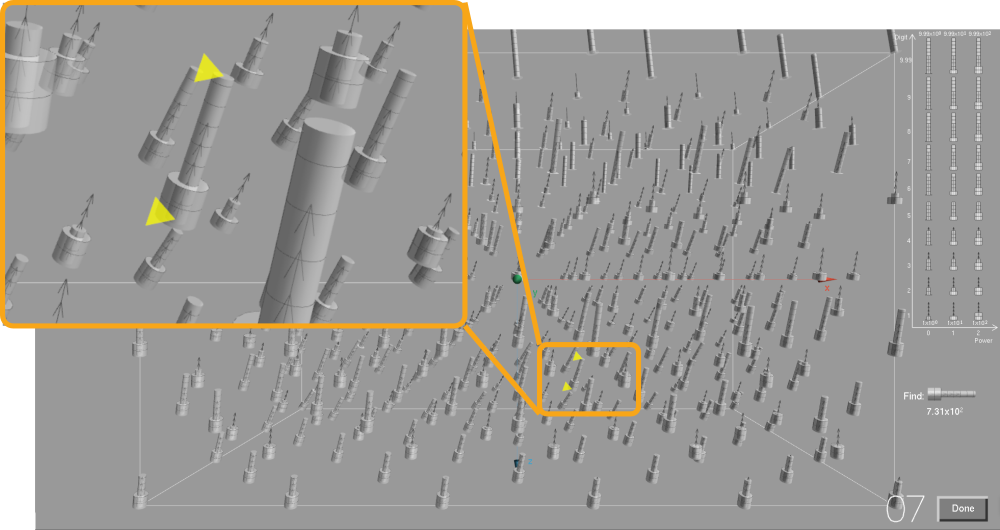}}; 
        		\end{tikzpicture}
        		\caption{SEARCH: Find the vector with magnitude $X$. ($X$: $731$, answer: the point marked by two yellow triangles.) }
        		\label{fig:SEARCH}
        	\end{subfigure}

			\begin{subfigure}[t]{0.45\textwidth}
        		\centering
        		\begin{tikzpicture}
        		\node (img1){\includegraphics[width=\textwidth]{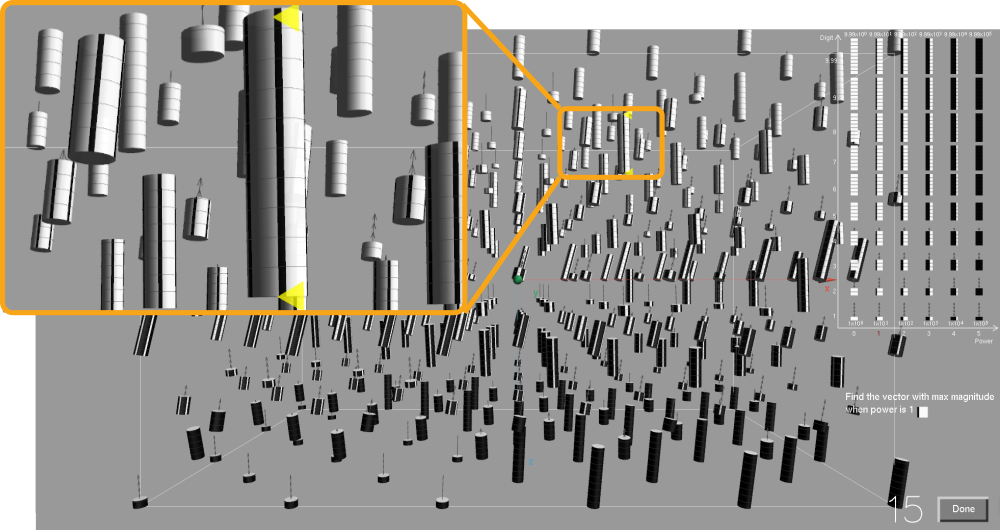}}; 
        		\end{tikzpicture}
        		\caption{MAX: Which point has the maximum magnitude when the exponent is $X$? ($X$: $1$, answer: the point marked by two yellow triangles.) }
        		\label{fig:MAX}
        	\end{subfigure}       

          \begin{subfigure}[t]{0.45\textwidth}
        		\centering
        		\begin{tikzpicture}
        		\node (img1){\includegraphics[width=\textwidth]{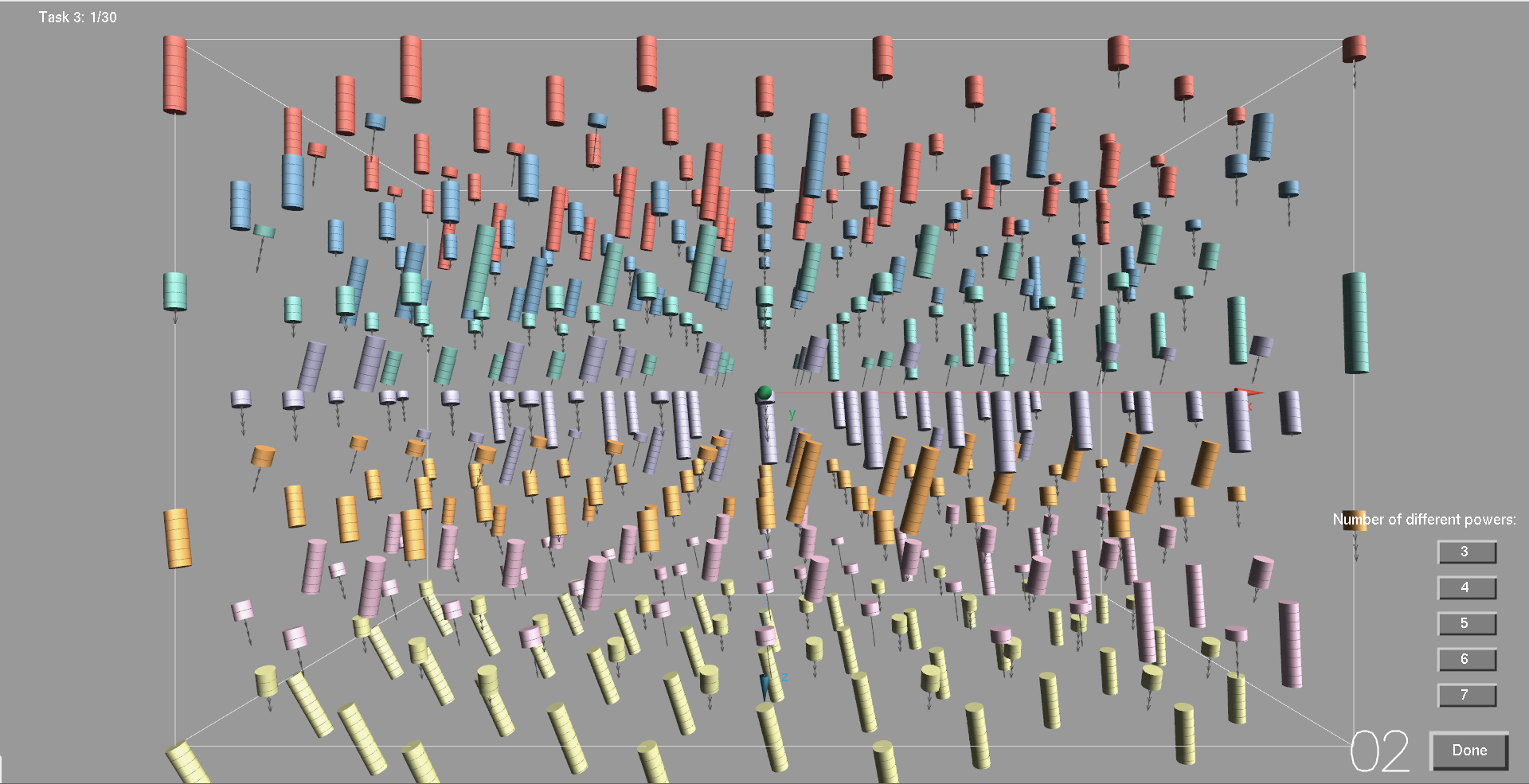}}; 
        		\end{tikzpicture}
        		\caption{NUMEROSITY (NUM): Estimate the total number of \changesr{unique} vector exponents of the entire vector field within $2$ seconds. (answer: $7$)}
        		\label{fig:NUM}
        	\end{subfigure}
        	
			 \caption{Experiment II three task types. The callouts show the task-relevant feature-pair(s).}
             
             \label{fig:exp2tasks}
        \end{figure}

\noindent Participants performed three tasks in which they had to compare all vectors to obtain an answer. 

\textbf{Exp II. Task 1 (SEARCH): visual search.}
	A vector search within $20$ seconds  (Figure~\ref{fig:SEARCH}).
\textit{Find the vector with magnitude $X$ within $20$ seconds.}
The target vector was shown at the bottom-right corner of the screen. Participants were asked to find this vector.

\textbf{Exp II. Task 2 (MAX): find maximum.}
 An  extreme value search within $20$ seconds (Figure{~\ref{fig:MAX}}).
\textit{Within $20$ seconds, locate the point \change{has}{of} maximum magnitude when the exponent is $X$.}
$X$ in the study was a number from $0$ to the \textit{maximum exponent} ($\in[2, 6]$). 
This was a global task requiring participants to find the extremum among many vectors.

\textbf{Exp II. Task 3 (NUMEROSITY): estimate the total number of \changesr{unique} vector exponents} (Figure~\ref{fig:NUM}). \textit{Estimate the total number of unique vector exponents in the entire vector field within $2$ seconds.}
Data are randomly chosen and modified to produce the $3$ to $7$ range. 

\subsubsection{Task Choices}

\changes{Tasks are \textit{use-inspired} 
by real-world quantum physics data analyses. 
Experiment I drilled down to a single or at most two spins. 
But global tasks are also of quantum physicists' interests, such as those involving understanding the distributions of quantum spin magnitudes. 
Practically, a spin
represents charge density or the measure of the probability of an electron being present at an infinitesimal element of space surrounding any given point. 
This probability varies due to electron traveling from one grid point to another and is often interpreted together with its neighbors.
Quantum physicists are thus interested in searches for regions,
where local regions are defined by spin magnitude 
and different regions would correspond to changes in exponent.
Often the most interesting regions are also those with specific charge densities (Task 1) or largest magnitudes (Task 2) .
The regional task is related to learning the number of interesting regions or magnitude exponent clusters (Task 3).
}



\changes{Performing tasks was limited to $20$ seconds as 
a pilot study 
showed that it took participants about $\in [15, 25]$ seconds or on average about $20$ seconds to finish search tasks 1 and 
2.
Also, preattentive processing when used for scene guidance involving a group of similar objects are often fast for viewers to see
and increasing the number of items should not significantly impair the search time.
From the practical side for the last \change{task}{experiment}, participants who would want a perfect score could just spend time counting. Constraining the time 
allowed us to measure the accuracy when they may have to use the scene feature. 
}

\subsubsection{Data Choices}

Data were  first sampled using  the same  approach as Experiment I, and no data is used repeatedly in this experiment. We then modified the exponent range from $3$ to $7$ for the three  tasks by normalizing the data  to the desired new  data  range. 

Prior  literature used  both  synthetic data and  real-world data  to  construct the  data  visualization as  test  scenarios, enabling tight  control  over  the  stimulus parameters (e.g., ~\cite{forsberg2009comparing}). Most of  the  synthetic data in literature were to replicate real-world data  characteristics; and others were explained in fictitious use scenarios.  
The goal was primarily to prevent preconceived user knowledge about the  domain-specific attributes. 
As a result, the synthetic data  strike  the  right balance  between real-world uses  and the  data  characteristics.

In our  cases,  replicating characteristics in quantum physics  data  was  challenging and indeed  impossible, since atom  behaviors in  high-dimensional space  were  largely  unknown and thus  were  not  easily  simulated. Our approach was  therefore to  randomly sample quantum physics  simulation results to capture domain-specific  attributes and  then  modify the  data  to  suit  evaluation purposes. We  showed our  data  to  our physicist collaborators to ensure their validity.
\changesr{We confirmed that these modifications preserved the  domain-specific schema of a scene in terms of the domain-specific structures and complexity from real simulations. These modifications represented less than $4\%$ of overall data points in each scene. Finally,  It improves the reuse  of our study results.
}



\subsubsection{Empirical Study Design}
\emph{Dependent and Independent Variables.}
We used a within-subject design with two  independent  variables of 
\textit{feature-pair} (three levels: baseline \textit{splitVectors}, \lengthycolor, and \lengthytexture) and \textit{exponent range} (five levels: $3-7$). 
The dependent variable was  relative error.  We did  not measure time since all tasks  were  time-constrained.

Participants performed $3$ (feature-pairs) $\times$ $5$ (magnitude-ranges) \add{$\times$ $3$ (repetitions)} $=$ \change{$15$}{$45$} trials for the first two tasks. Three repetitions were used to give participants enough time to develop strategies. 
For NUMEROSITY tasks, the design runs $4$ repetitions, resulting in $3$ (feature-pair\add{s}) $\times$ $5$ (exponent-range\add{s}) $\times$ $4$ (repetition\add{s}) $=$ $60$ trials. Each participant thus executed $45+45+60$ $=$ $150$ trials. Completing all tasks took about $32$ minutes.

\emph{Self-Reporting Strategies.} 
Several  human-computer interaction   (HCI)  approaches can  help   observe users'  behaviors.  Answering questions can  assist  us  to  determine not just  which  technique is better  but  also  the strategies humans adopt. For  example, cognitive walkthrough  (CTW)  measures  whether or  not  the  users'  actions  match  the  designers'  pre-designed  steps.   Here   we  predicted  that   participants would use  the  global  scene-features as  guidance to accomplish tasks. We interviewed participants and  asked them  to verbalize their  visual  observations in accomplishing tasks.

\subsubsection{Participants}
Eighteen new  participants ($12$ male and $6$ female, mean age = $23.8$, and standard deviation = $4.94$) of diverse backgrounds participated in the  study (seven in computer science, four in computer engineering, two in information systems, three in engineering, one in business school, and one in physics).

Procedure, interaction, and  environment were  the same  as those in the Experiment I.

\subsection{Experiment II: Results and Discussion}

We collected $810$ data points per task for the first two tasks of SEARCH and MAX and $1080$ points for the third NUMEROSITY task. 

\subsubsection{Analysis Approaches}

For SEARCH and MAX tasks, we measured relative error (which was the percentage the reported value was away from the ground truth and the same as that of Experiment I) with SAS repeated measure. \remove{for SEARCH and MAX}
The last NUMEROSITY task used 
error rate
which was the percentage of 
\changesr{incorrect} 
answers of all trials for each participant. 
We also used the same outlier removal methods to remove instances of correspondence errors for SEARCH and MAX.


\subsubsection{Overview of Study Results}

\begin{figure}[!tbp]
\centering
	\begin{subfigure}[t]{0.48\textwidth} 
        	\includegraphics[width=\textwidth]{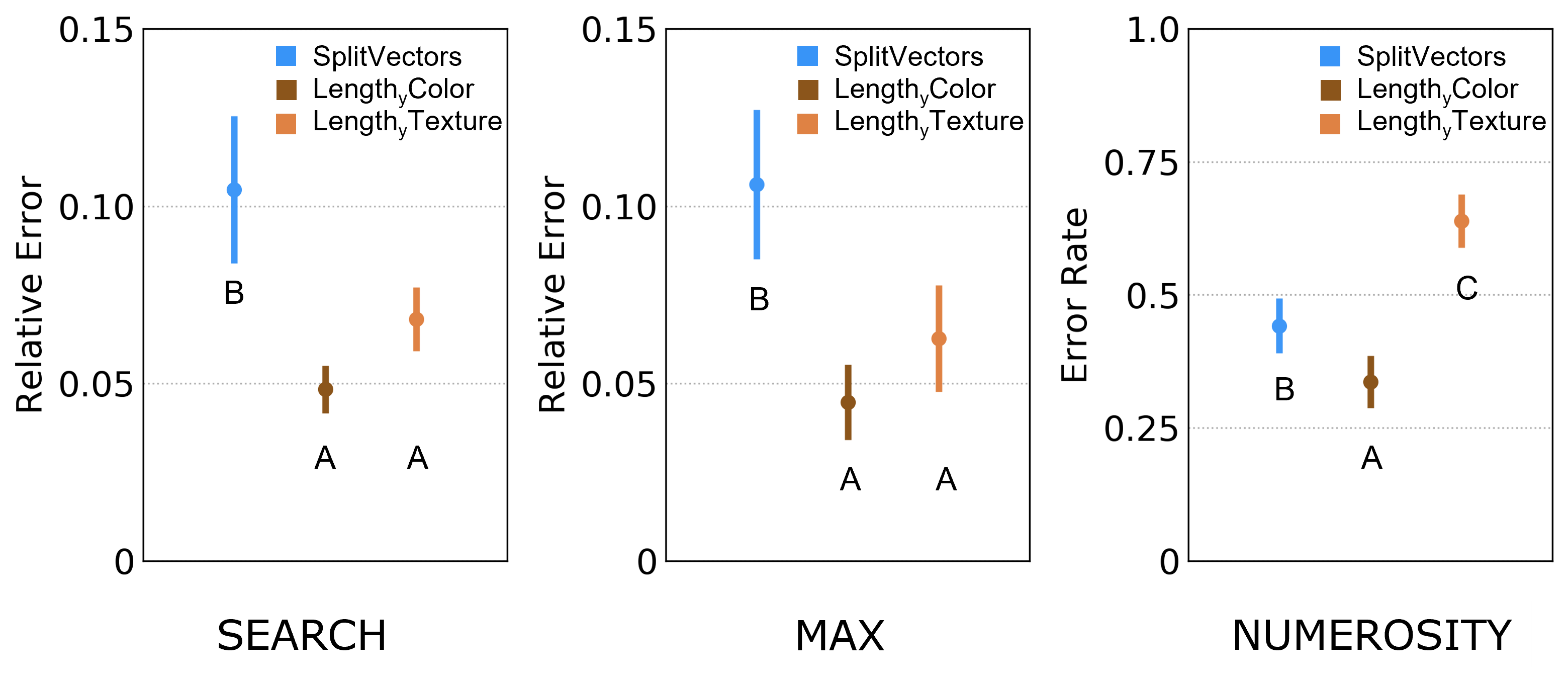}
		\label{fig:exp2max_feature}
	\end{subfigure}
	\begin{subfigure}[t]{0.48\textwidth}
        	\includegraphics[width=\textwidth]{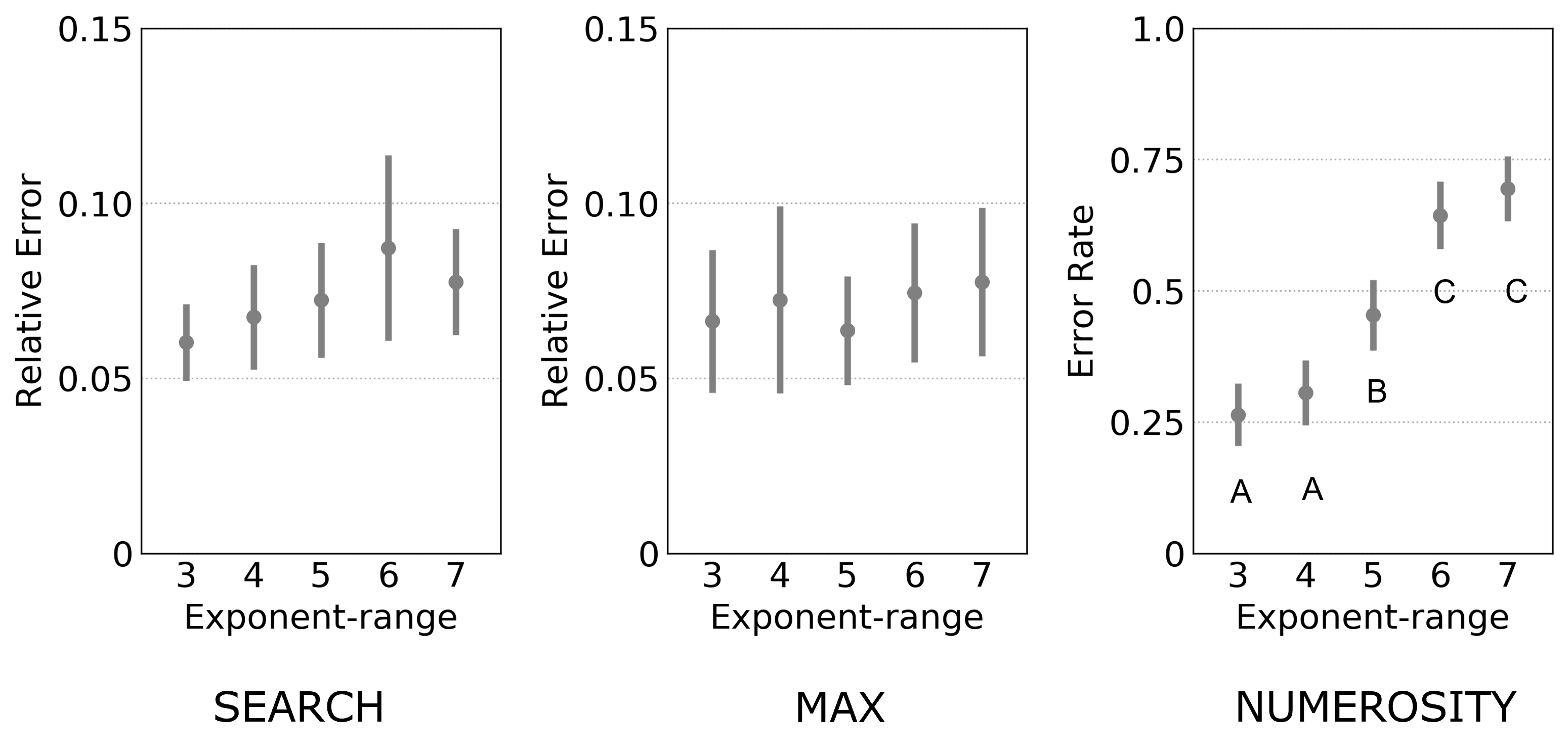}
		\label{fig:exp2num_feature}
	\end{subfigure}
\caption{\changes{Relative error for Tasks SEARCH and MAX  was the percentage the reported value was away from the ground truth. 
Error rate for NUMEROSITY \remove{which} was the percentage of 
wrong
answers of all trials for each participant.
The vertical axis shows the relative error or error rate.}
Same letters represent the same post-hoc analysis group. All error bars represent $95\%$ confidence intervals.
}
\label{fig:exp2result}
\end{figure}

Table~\ref{tab:exp2result18} and Figure~\ref{fig:exp2result} show the summary statistics; And all error bars again represent $95\%$ confidence intervals.
We observed a significant main effect of feature-pair type on all three tasks. 
For the first two tasks, the post-hoc analysis revealed that \lengthycolor and \lengthytexture were in the same group, the most efficient one and that relative errors were statistically significantly lower than those of the \textit{splitVectors}. \Lengthycolor remained the most accurate pair for the NUMEROSITY tasks. 
Exponent-range was only a significant main effect for NUMEROSITY, with power ranges $3$ and $4$ were significantly better than $5$, which was better than $6$ and $7$.

\subsubsection{More Categorical Features of Separable Dimensions Improved Accuracy}

We  were  interested to see if we  could   observe significant main effects of categorical features in the separable pairs in this experiment. 
Here  we did observe the significant main effect and confirmed our first hypothesis (H1) for both SEARCH and MAX: in the general trend, more separable \lengthycolor was more effective than \lengthytexture which was better than \textit{splitVectors}, and \lengthycolor  and  \lengthytexture were  in  the  same Tukey group,
\changes{when viewers \add{were} in the correct data sub-categories.} 

\Lengthycolor led  to  the most  accurate answers, and \textit{splitVectors} was  better  than   \lengthytexture for NUMEROSITY task. This result can be explained by participants' behaviors - more  than half the  participants suggested they simply look for the longest cylinder from \textit{splitVectors} since they know  the numerical values in the test were continuous. This behavior deviated from  our  original purpose of testing the global estimate but did show  two perspectives in favor of this  work:  (1) participants developed task-specific strategies during the experiment for efficiency; 
\changesr{(2) 3D length still supported judging large and small and it was not as effective as color perhaps due to ensemble perception from categorical features.
}

\begin{figure}[!tbp]
\centering
        	\includegraphics[width=0.85\columnwidth]{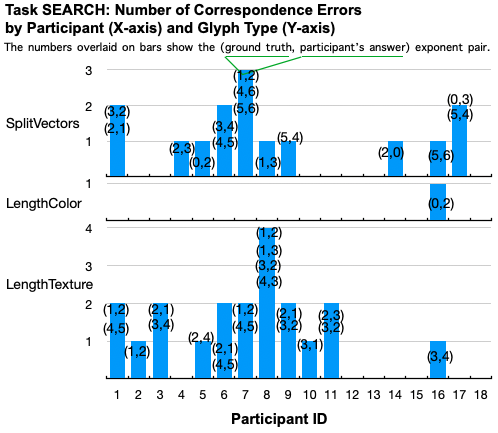}
		\label{fig:exp2searchce}
        	\includegraphics[width=0.85\columnwidth]{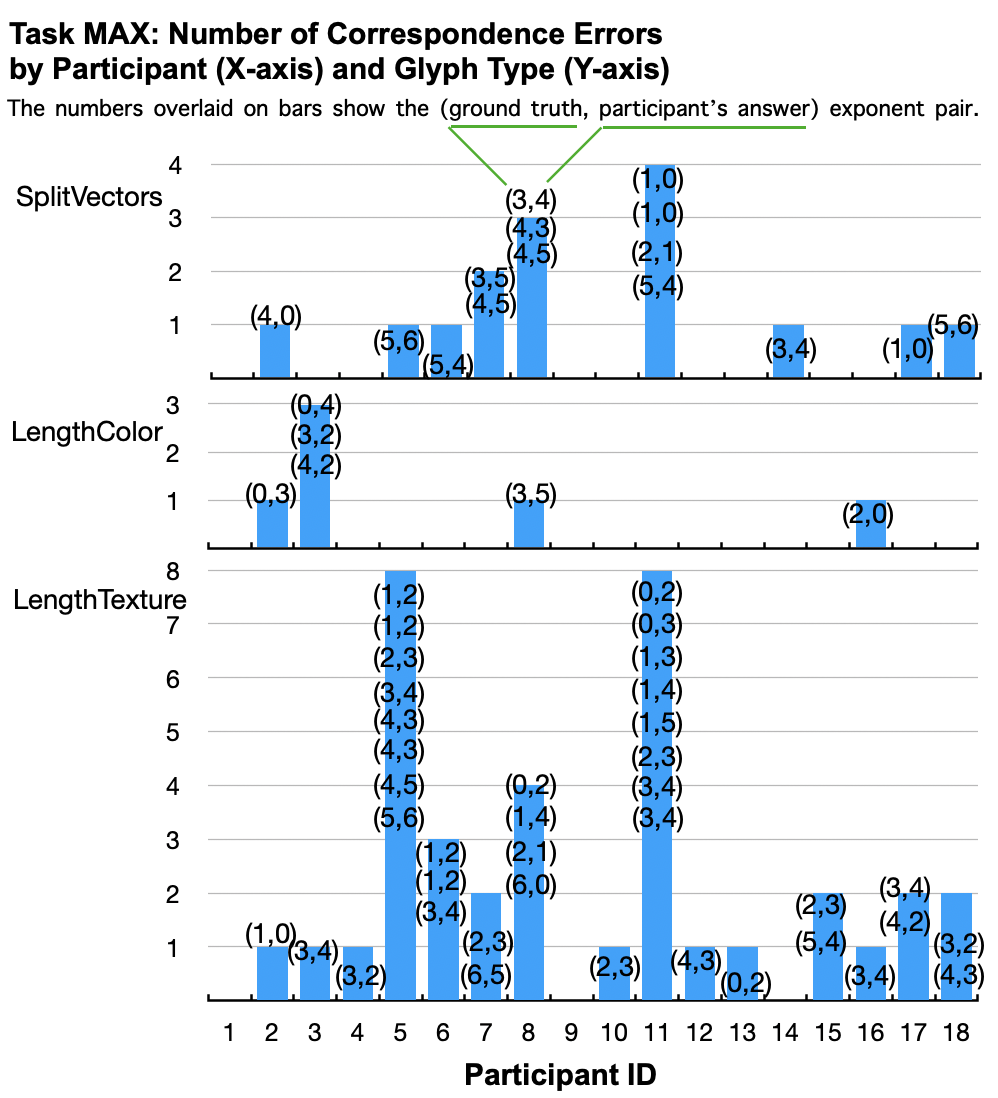}
		\label{fig:exp2maxce}
\caption{\changes{Experiment II (Tasks SEARCH and MAX): All instances of correspondence errors by participant. Again, the \lengthycolor has the least instances of correspondence error whilst the \lengthytexture had the most.}
}
\label{fig:exp2ce}
\end{figure}

\subsubsection{Color Categories  of Separable Pairs Reduced Correspondence Errors by a Large Margin}

Our second hypothesis H2 was also supported. We  first tested the number of 
\changesr{correspondence errors} 
in SEARCH and MAX in the same way as in Experiment I.
These results when combined with those in Experiment I confirmed again that the \lengthycolor reduced correspondence errors. 
\changesr{For SEARCH, There were only a single instance of correspondence error.}
$36$ instances of correspondence errors came
from $14$  participants (mean$=2.57$, $95\%$ CIs=$[2.1, 3.04]$) (Figure~\ref{fig:exp2ce} \textit{top}).
Another $59$ instances for MAX came from $16$ of $18$  participants, 
mean$=3.68$, $95\%$ CIs$=[2.85, 4.51]$) (Figure~\ref{fig:exp2ce} \textit{bottom}).

\begin{table}[!tp]
	\caption{Exp II: Summary statistics by tasks. The significant main effects and the high effect size are in \textbf{bold} and the medium effect size is in \textit{italic}. Effect size is 
	Cohen's d for tasks SEARCH and MAX, and Cramer's V for task NUMEROSITY (NUM). Post-hoc Tukey grouping results are reported for significant main effects, where $>$ means statistically significantly better and enclosing parentheses mean they belong to the same Tukey group. Here, \lc: \lengthycolor and \lt: \lengthytexture.}
	\label{tab:exp2result18}
	\vspace{-4mm}  
	\begin{center}
		\begin{tabular}{l l l l}
		\toprule
\centering Task & Variables & Significance & ES\\
			\midrule
    	SEARCH & feature-pair & \textbf{F$_{(2,\,261)}$ = 18.4}, \textbf{\textit{p} $<$ 0.0001} & \textbf{0.46}\\
    	& & \textbf{(\lc, \lt)} $>$ \textbf{\textit{splitVectors}}& \\
		& power-range & F$_{(4,\,261)}$ = 3.0, p = 0.20  & \textbf{0.86}\\
			\midrule	
	MAX & feature-pair & \textbf{F$_{(2,\,261)}$ = \change{15.8}{15.4}}, \textbf{\textit{p} $<$ 0.0001} & \textbf{0.47} \\ 
	& & \textbf{(\lc, \lt)} $>$ \textbf{\textit{splitVectors}}& \\
		& power-range & F$_{(4,\,261)}$ = 0.3, \textit{p} = 0.87 & 0.11\\
			\midrule
	NUM & feature-pair &  \textbf{${\chi}^2$ = 63.2}, \textbf{\textit{p} $<$ 0.0001} & \textit{0.25} \\
	& & \textbf{\lc} $>$ \textbf{\textit{splitVectors}} $>$ \textbf{\lt}& \\
   & power-range &  \textbf{${\chi}^2$ = 47.4}, \textbf{\textit{p} $<$ 0.0001}  &  \textbf{0.35} \\
   & &  \textbf{(3, 4)} $>$ \textbf{5} $>$ \textbf{(6, 7)}&\\
			\bottomrule
		\end{tabular}
	\end{center}
\end{table}

\subsubsection{\changes{Compensating The Cost of Search in Complex Data through Preattentive Scene Feature}}
\changes{
The visualizations in our study 
contained hundreds of items from realistic uses.}
Subjective behaviors through self-report suggested that they
adopted a
sequential task-driven viewing strategy to first obtain gross regional distribution
of task-relevant exponents. After this, a visual comparison within the same exponent region were achieved.
With these two steps, judging large or small or perceiving quantities accurately from separable variables  would not use object-level information process.

Many participants commented on how  the number of powers in the data affected  their effectiveness. For  \lengthytexture, $10$ participants remarked that  it was  difficult to differentiate adjacent powers when the total power level is around $4$-$5$ for \lengthytexture. The  white  and  black textures  were   very  easy  to  perceive. All but two participants agreed that  \lengthycolor  could  perhaps support up  to  $6$.  Chung et al.{~\cite{chung2016ordered}} studied ordering effects and it would be challenging to compare ours to their results because their visual features were not shown as a scene but \change{a stimuli alone}{an isolated feature}. 
More  than  half  of  the participants felt  that  effectiveness  of \lengthylengthy 
was not affected  by  changing the  number of powers, since they  looked  for  the  longest outer cylinder to  help  find  the answer.
These results may suggest that subregion selection with \lengthytexture can perhaps be better designed with interfaces when the users can interactively select a texture level. 

\section{General Discussion}

We discuss the results from both experiments and suggest future directions.

\subsection{\changes{Separable Dimensions with Preattentive Guidance for Large-Magnitude-Range Quantum Physics Spins}}

\changesr{Our first principle in glyph design is to follow the convention to use separable variable pairs~\cite{borgo2013glyph, ware2020information}.}
The  results  of  Experiment I  showed  that  separable dimensions could achieve the same efficiency as direct linear visualizations for COMP and was always more efficient than integral pairs. For these local-tasks, we didn't observe significant error reduction.

Our second principle in glyph design is to 
\changesr{include categorical features in separable pairs.}
The results from Experiment II studied the rank order of the separable pairs and found that they indeed  improved accuracy for global tasks. \Lengthytexture and \textit{splitVectors} in both experiments led to more 
\changes{correspondence errors} than \lengthycolor.
\changes{Achieving integrated numerical readings by combining two separable visual features at object level seems not necessary.}

The separable-dimension pairs of \lengthycolor and \lengthytexture worked because they were preattentive scene features. 
Our experiment\add{s} show that 
viewers adopted a sequential task-driven viewing strategy based on a view hierarchy: viewers first  obtain  \textit{global}  distributions of the scene. Then, a visual scrutiny is possible within a subregion. 
Although \textit{splitVectors} are separable, visual search for length among length would be unguided because both targets and distractors contained the same visual variable. The more separable, the easier it would be to guide the attention. 
Using coloring to provide some initial regional division may be always better than not. Texture (luminance) could achieve similar accuracy and efficiency as long as \changes{the task-relevant regions could be detected}.


\begin{figure*}[!tb]
\centering
\begin{subfigure}{0.48\textwidth}
		\includegraphics[width=\textwidth]{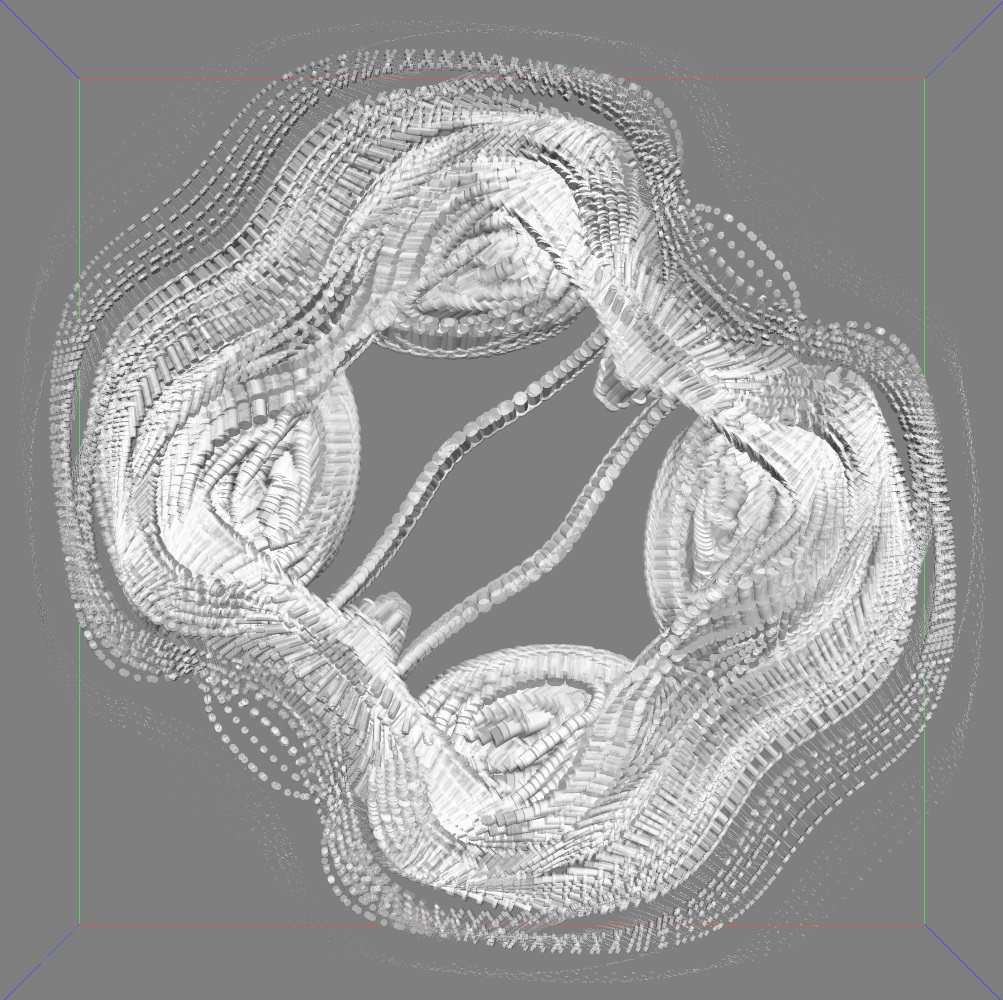}
		\caption{\Lengthylengthx feature-pair}
		\label{fig:case2LLO}
\end{subfigure} 
\begin{subfigure}{0.48\textwidth}
		\includegraphics[width=\textwidth]{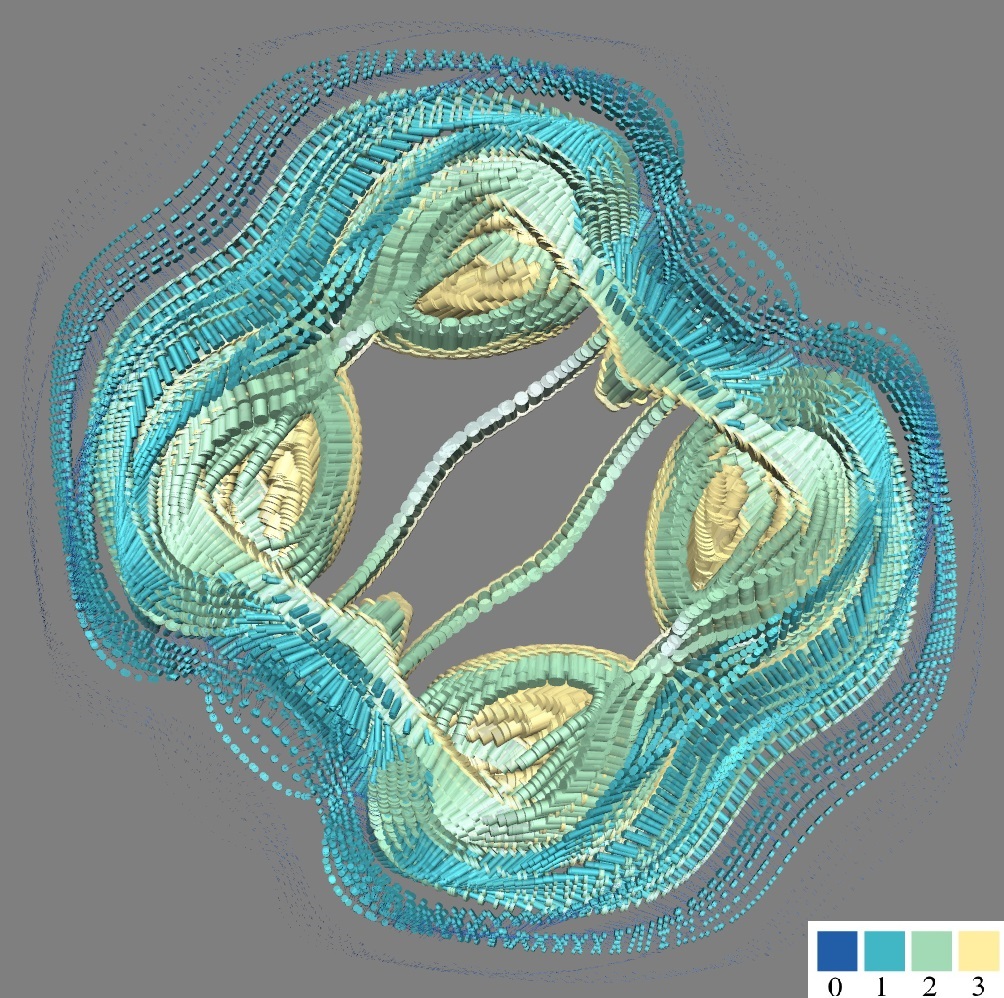}
		\caption{\Lengthycolorlengthx feature-pair}
		\label{fig:case2LCL}
\end{subfigure} 
\begin{subfigure}{0.48\textwidth}
		\includegraphics[width=\textwidth]{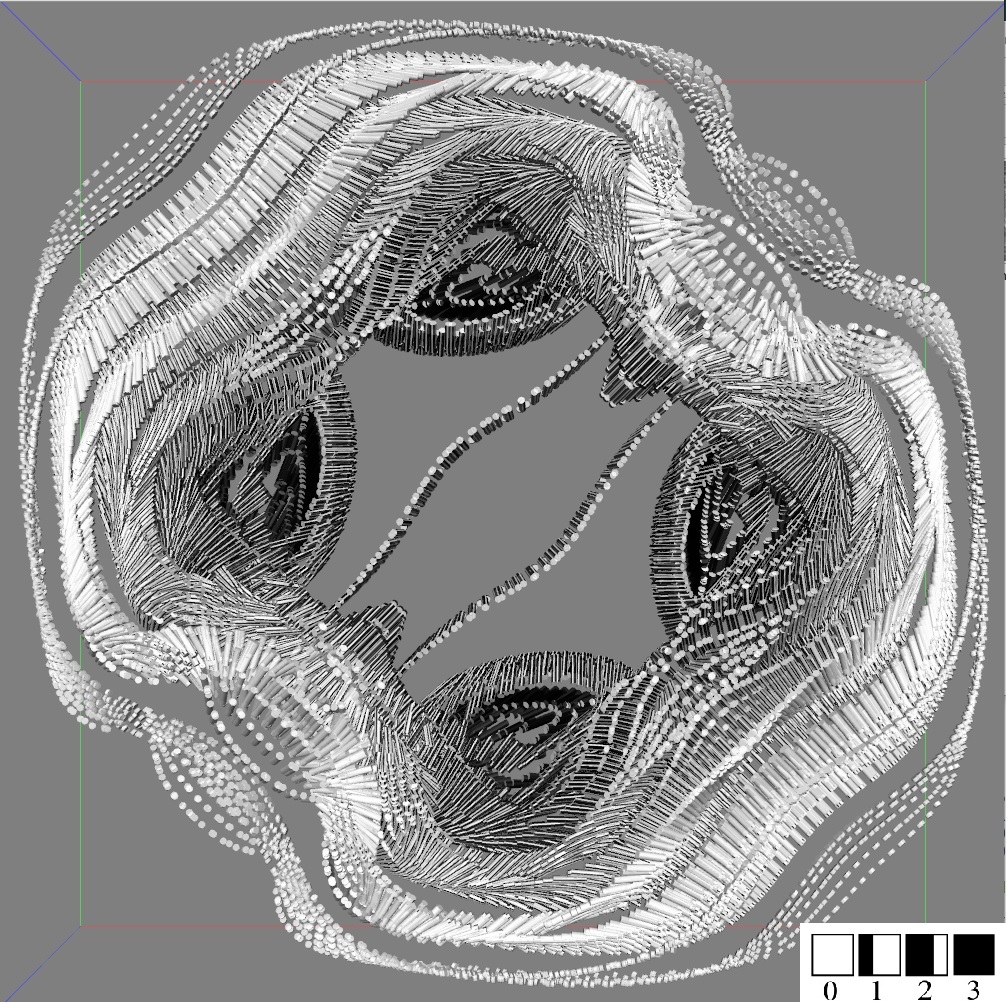}
		\caption{\Lengthytexture feature-pair}
		\label{fig:case2LT}
\end{subfigure} 
\begin{subfigure}{0.48\textwidth}
		\includegraphics[width=\textwidth]{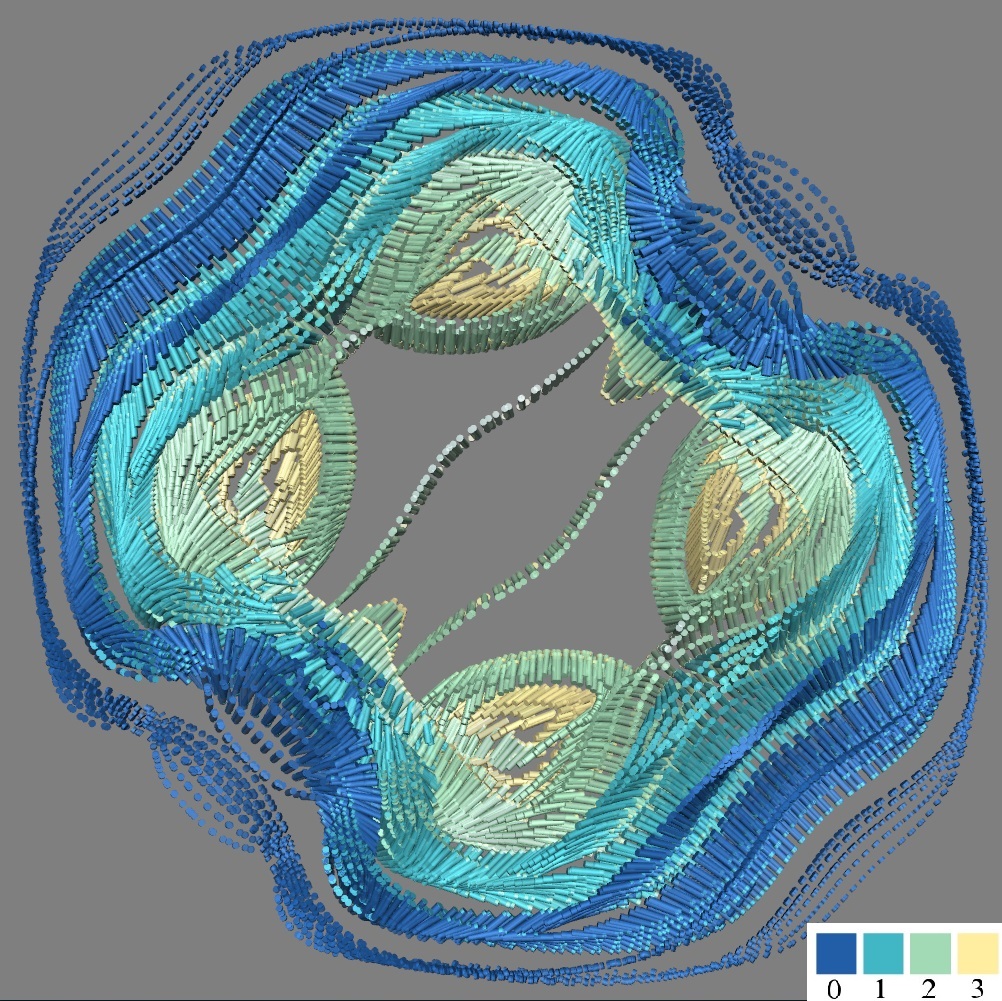}
		\caption{\Lengthycolor feature-pair}
		\label{fig:case2LC}
\end{subfigure} 
\caption{
Contours of simulation data. Size from this viewpoint can guide visual grouping and size in 3D must take advantage of knowledge of the layout of the scene~\cite{eckstein2017humans}. }
\label{fig:case2}
\end{figure*}

\subsection{Feature Guidance vs. Scene Guidance}

Taking into account both study results, we think an important part of the answer to 
visualization design
is \textit{guidance} of attention. 
It is guided to some objects or locations over others by two broad methods: \textit{feature guidance (seeing objects)} and \textit{scene guidance} (seeing global structures).  

Feature guidance refers to guidance by properties of the task-target as well as the distractors (leading to correspondence errors). These features are limited to a relatively small subset of visual dimensions: color, size, texture,  orientation, shape, blur or shininess and so on.
These features have been broadly studied in 3D glyph  design (see reviews by Healey  and  Enns~\cite{healey2012attention}, Borgo  et al.~\cite{borgo2013glyph}, Lie et al.~\cite{lie2009critical},  Ropinski et al.~\cite{ropinski2011survey}, and McNabb and Laramee~\cite{mcnabb2017survey}). 
Take one more example from quantum physics simulation results, but with a different task of searching for the structural distributions  in the power of $3$ in Figure~\ref{fig:case2} will guide attention to either the fat cylinders (Figure~\ref{fig:case2LLO}) or the bright yellow color (Figure~\ref{fig:case2LC}, ~\ref{fig:case2LCL}) or the very dark texture (Figure~\ref{fig:case2LT}), depending on the feature-pair types. 

Working with quantum physicists, we have noticed that
the \textit{structure and content of the scene} strongly constrain the
possible location of meaningful structures, guided ``scene
guidance'' constraints~\cite{biederman1977processing, wolfe2021guided}. Scientific data are not random and are typically structured. Contextual and global structural influences can arise from different sources of
visual information. If we return to the MAX search task in
Figure~\ref{fig:case2} again, we will note that the chunk of darker or
lighter texture patterns and colors on these regular contour
structures strongly influence our quick detection. This is
a structural and physical constraint that can be utilized
effectively by viewers. This observation coupled with the
empirical study results may suggest an interesting future work and hypothesis: \textbf{adding scene structure guidance
would speed up quantitative discrimination, improve the
accuracy of comparison tasks, and reduce the perceived
data complexity}.

Another structure acting as guidance is the size itself. It was used by participants seeking to resolve the NUMEROSTIY tasks to look for the longest outside cylinders. We have showed several examples like Figure~\ref{fig:case2}, our collaborator suggested that the cylinder-bases of the same size with the redundant encoding (Figure~\ref{fig:case2}b) also helped locate and group glyphs belonging to the same magnitude. This observation agrees with the most recent literature that guidance-by-size in 3D must take advantage of knowledge of the layout of the scene~\cite{eckstein2017humans}. 

Though feature guidance can be preattentive and features are detected within a fraction of a second, scene guidance is probably just about as fast (though precise experiments have not been done and our Experiment II only
merely shows this effect). Scene `gist' can be extracted from
complex images after very brief exposures~\cite{biederman1977processing}~\cite{oliva2005gist}. This doesn’t mean that a viewer instantly knows, say, where the answer is located. However, with a fraction of a second's exposure, a viewer will know enough about the spatial layout of the scene to guide his or her attention towards
vector groups in the regions of interest. 
\changesr{For example, categorical color becomes scene features since these colorful glyphs were perceived as a whole}

A future direction, and  also an  approach to understanding the  efficiency  and  the  effectiveness of scene  guidance, is to conduct an eye-tracking study to give viewers a flash-view of our spatial structures and  then  let  the  viewer  see  the display only in a narrow range  around the point  of fixation: 
\textit{does  this  brief  preview guide attention and  the gaze effectively?} 
Work in vision and visualization{~\cite{ryan2019glance, bylinskii2015intrinsic, borkin2013makes, li2018toward}} domain has  measured and  correlated performance on  the glance  or  global  structure formation. Vision  science  discovered long ago that  seeing global  scene structures in medical imaging decision making guides experts’ attention (experts always know  where to look){~\cite{kundel2007holistic}~\cite{drew2013invisible}}.

\subsection{\changes{Redundancy and Ensemble Graphical Perception}}

\changesr{
Our results showed that adding categorical colors, in which the correspondence parts could be quickly discriminated, is scalable to a large number of items. }
Our result agrees with that of Northelfer and Gleicher~\cite{nothelfer2017redundant}. They observed that redundant encoding using color and shape could strengthen grouping when searching for targets from multiple objects. Their explanation was a race model~\cite{nothelfer2017redundant}: for separable dimensions, the performance of a glyph with the redundant encoding might be dominated by the feature with greater efficiency. We did not 
find efficiency improvement - this suggested that the grouping is generally fast. So it might \textit{not} be the redundancy itself that contributed to scene understanding.

\changes{Another possible theory is perhaps
\textit{ensemble perception}, i.e.,
``the visual system's ability to extract summary statistical information from groups of similar objects - often in a brief glance''~\cite{whitney2018ensemble}. Also ensemble features are best represented
using the categorical features.
To model parallel processing, the target contrast signal theory by Buetti et al.~\cite{buetti2019predicting} may suit our scenario better. It describes more specific time estimate it takes to evaluate items in \textit{parallel}. In visualization, we just began to understand the ensemble averages (e.g., Chen~\cite{chen2021ensemble} and Alberts et al.~\cite{albers2014task}) but have limited understanding of ensemble visual encoding choices to guide attention to optimize behaviors. 
We leave this to future work.
}

\subsection{Use Our Results in Visualization Tools and Limitations of Our Work}

Visualization is used when the goal is to augment human capabilities in situations where the problems might not be sufficiently defined for 
algorithms to communicate certain information. One of our showcase areas is quantum physics.
We believe that the design principle of prompting 
\changesr{the addition of categorical features in bivariate glyphs
would be broadly applicable to glyph design.}
Also, application domains carrying similar data attributes could reuse of work.
Our current study concerns bivariate data visualization in which the bivariate variables are component parts of scalar variables. 

Our design could have been improved by following advanced tensor glyph design methods.
Both generic{~\cite{gerrits2017glyphs}}  and domain-specific requirements for glyph designs {~\cite{zhang2016glyph}~\cite{schulz2013design}~\cite{kindlmann2006diffusion}} have led to the summary of glyph properties (e.g.,~invariant, uniqueness, continuity) to guide design and to render 2D and 3D tensors. A logic step 
is to truly understand the quantum physics principles to combine data attributes and human perception to improve our domain-specific 
solutions.  

One limitation of this work is that we measure only a subset of tasks crucial to showing structures and omitted all
tasks relevant to orientation. However, one may argue that the vectors naturally encode orientation. 
When orientation is considered, we could address the multiple-channel mappings in two ways.
The first solution is to use the \lengthytexture to encode the
quantitative glyphs and color to encode the orientation clusters.
The second solution is to treat magnitude and orientation as two
data facets and use multiple views to display them separately, with one view showing magnitude and the
other for orientation (using Munzner's multiform design recommendations~\cite{munzner2014book}).
The second limitation here was that our experiments were limited to  a relatively small  subset  of visual  dimensions: color, texture, and size. A future direction would be to try shapes and glyphs to produce novel and useful design.

\section{Conclusion}

\changes{
Our findings in general suggest that, as we hypothesized, distinguishable separable dimensions with preattentive categorical features perform better. 
The separable pair \lengthycolor
was the most efficient and effective for both local and global tasks. \changesr{The categorical features enable effective complex scene inspections.}
Our empirical study  results provide the following recommendations for  designing 3D bivariate glyphs.
}.

\begin{itemize}
\item
Highly separable pairs can be used for quantitative 
\changesr{comparisons as long as these glyphs could guide attention (i.e., category forming). 
We recommend using \lengthycolor.}

\item
Texture-based glyphs (\lengthytexture) that introduces {luminance} variation
will only be recommended when task-relevant structures can be \changes{isolated}. 

\item
Integral and separable bivariate {feature-pairs} have the similar accuracy when the tasks are {local.} 
\changesr{
When the search tasks are more complex, introducing categorical features in the separable feature-pairs will lead to perceptually accurate glyphs.
}

\item
\changesr{3D glyph scene would shorten task completion time by combing two glyph design factors: separability and visual guidance from categorical features.}

\item
The redundant encoding (\lengthycolorlengthx) greatly improved on 
{task completion time} of integral dimensions (\lengthylengthx) by adding separable and preattentive color features.
\end{itemize}

\ifCLASSOPTIONcompsoc
  \section*{Acknowledgments}
  The work is supported in part by NSF IIS-1302755, NSF CNS-1531491, and NIST-70NANB13H181.
The user study was funded by NSF grants with
the OSU IRB approval number 2018B0080. 
Non-User Study design work was supported by grant from NIST-70NANB13H181. 
The authors would like to thank Katrina Avery for her excellent editorial support 
and all participants for their time and contributions. 

Any opinions, findings, and conclusions or recommendations expressed in this material are those of the author(s)
and do not necessarily reflect the views of the National Science Foundation. Certain commercial products are
identified in this paper in order to specify the experimental procedure adequately. Such identification is not
intended to imply recommendation or endorsement by the National Institute of Standards and Technology, nor
is it intended to imply that the products identified are necessarily the best available for the purpose.
\else
  \section*{Acknowledgment}
\fi

\ifCLASSOPTIONcaptionsoff
  \newpage
\fi

\bibliographystyle{IEEEtran}
\bibliography{ms}
\vspace*{-60pt}

\begin{IEEEbiography}[{\includegraphics[width=1in,height=1.25in,clip,keepaspectratio]{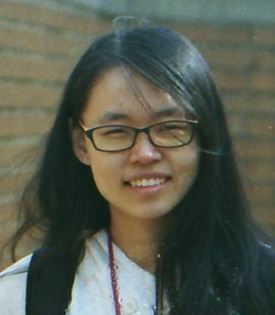}}]{Henan Zhao}
	was a PhD student
	in Department of Computer Science and Electrical Engineering at University of Maryland, Baltimore County. She received B.E. degree in Computer Science and Information Security from Nankai University, China. Her research interests include design and evaluation of perceptually accurate visualization techniques. This work was conducted while she was visiting The Ohio State University.
\end{IEEEbiography}
\vspace*{-50pt}



\begin{IEEEbiography}[{\includegraphics[width=1in,height=1.25in,clip,keepaspectratio]{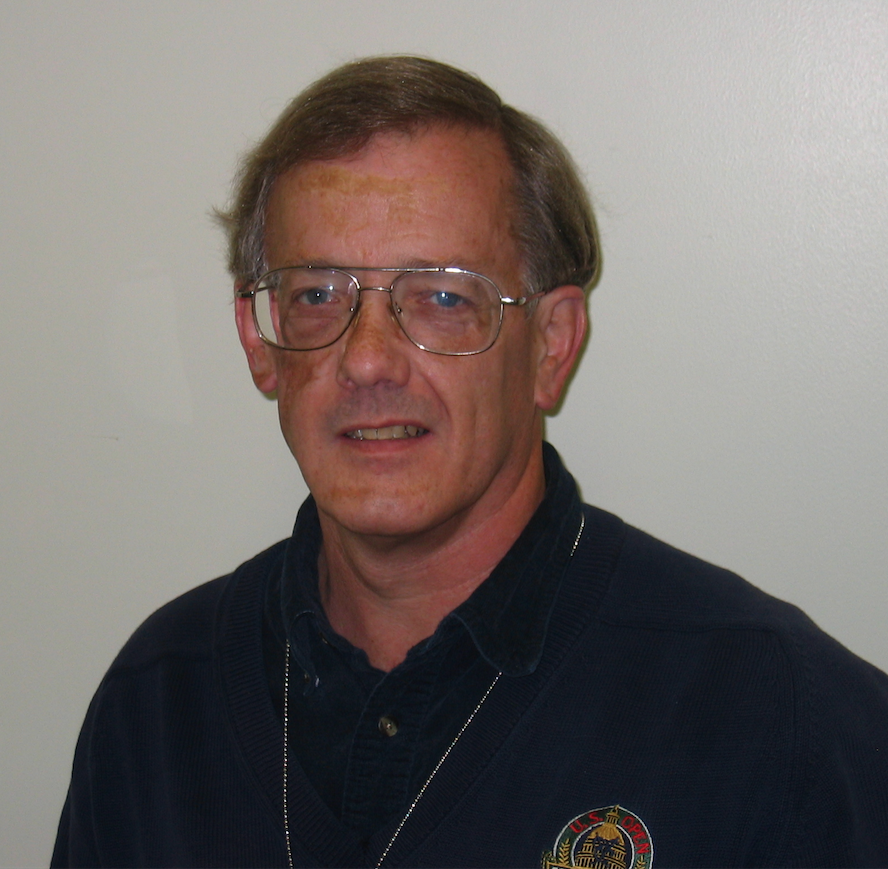}}]{Garnett Bryant}
	received his PhD at Indiana University in theoretical condensed matter physics. After research positions at Washington State University, the National Bureau of Standards, McDonnell Research Labs and the Army Research Laboratory, he has worked at the National Institute of Standards 
	and Technology (NIST) since 1994. He is directing the Quantum Processes and Metrology 
	Group at NIST with experimental and theoretical programs on nanoscale, condensed matter systems for 
	quantum information science and metrology. He is a Fellow of the Joint Quantum Institute of 
	NIST/University of Maryland, a Fellow of the American Physical Society and a member of the IEEE. His 
	theoretical research program focuses on nanosystems, nanooptics and quantum science.
\end{IEEEbiography}

\vspace*{-50pt}

\begin{IEEEbiography}[{\includegraphics[width=1in,height=1.25in,clip,keepaspectratio]{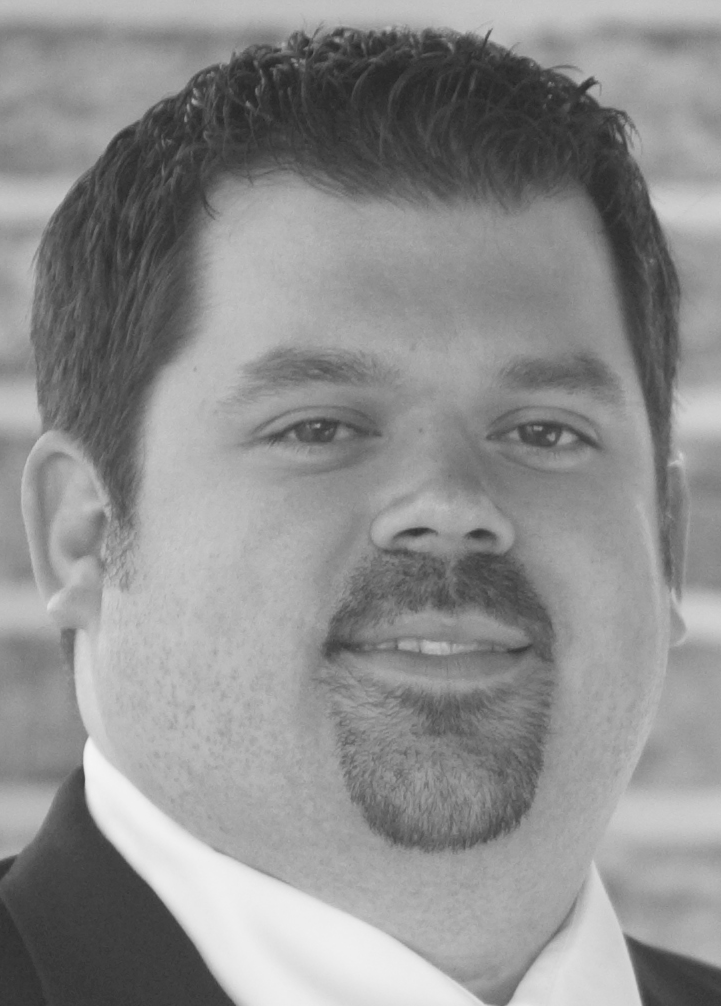}}]{Wesley Griffin}
	received his PhD degree in Computer Science from the
	University of Maryland, Baltimore County. He is a developer at Stellar Science. 
	His research
	interests include real-time graphics and graphics hardware. He is a
	member of ACM SIGGRAPH, the IEEE and the IEEE Computer Society.
\end{IEEEbiography}

\vspace{-50pt}

\begin{IEEEbiography}[{\includegraphics[width=1in,height=1.25in,clip,keepaspectratio]{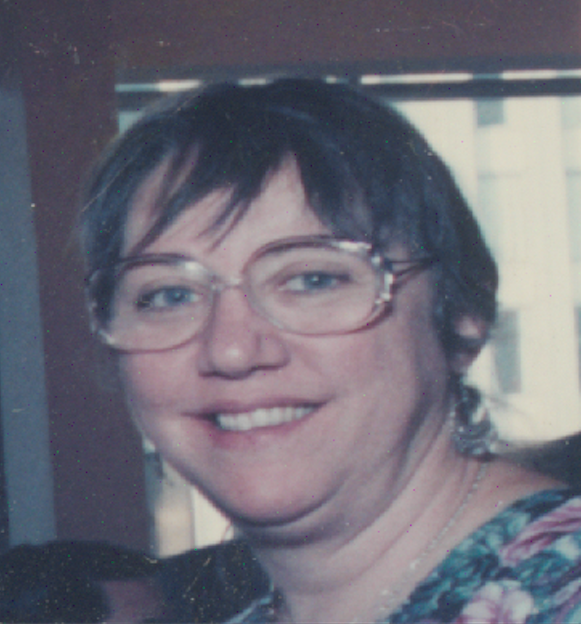}}]{Judith E. Terrill}
	is a Computer Scientist and the Leader of the High Performance Computing and Visualization Group at the National Institute of Standards and Technology. She is a member of the IEEE Computer Society, the Association for Computing Machinery, and the Association for the Advancement of Artificial Intelligence.
\end{IEEEbiography}

\vspace*{-50pt}

\begin{IEEEbiography}[{\includegraphics[width=1in,height=1.25in,clip,keepaspectratio]{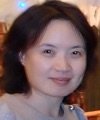}}]{Jian Chen} 
is an Associate Professor in Computer Science and Engineering at The Ohio State University. She received her PhD degree in Computer Science from Virginia Tech,
and her MS degree in Mechanical Engineering $|$ Precision Instrument from Tianjin University $|$ Tsinghua University, China. She was a
postdoctoral fellow at Brown University and a visiting researcher at Harvard University. Her current research interests include visual design, 3D interaction, and human-AI teaming.
\end{IEEEbiography}
\vspace*{-50pt}





\clearpage
\onecolumn
\noindent\begin{minipage}{\textwidth}
	\vspace{1cm}
	\makeatletter
	\centering
	\sffamily\LARGE\bfseries
	\@title\\[1em]
	\large Additional Material\\[1em]
	\makeatother
\end{minipage}
\vspace{1cm}

	

\noindent Empirical study training documents, source code, study data, and results are online at \href{https://osf.io/4xcf5/?view_only=94123139df9c4ac984a1e0df811cd580}{$https://osf.io/4xcf5/?view_only=94123139df9c4ac984a1e0df811cd580$}.

\section*{A. Background Color}
\label{sec:appendixb}

\begin{figure*}[!bp]
  \centering
\includegraphics[width=\linewidth]{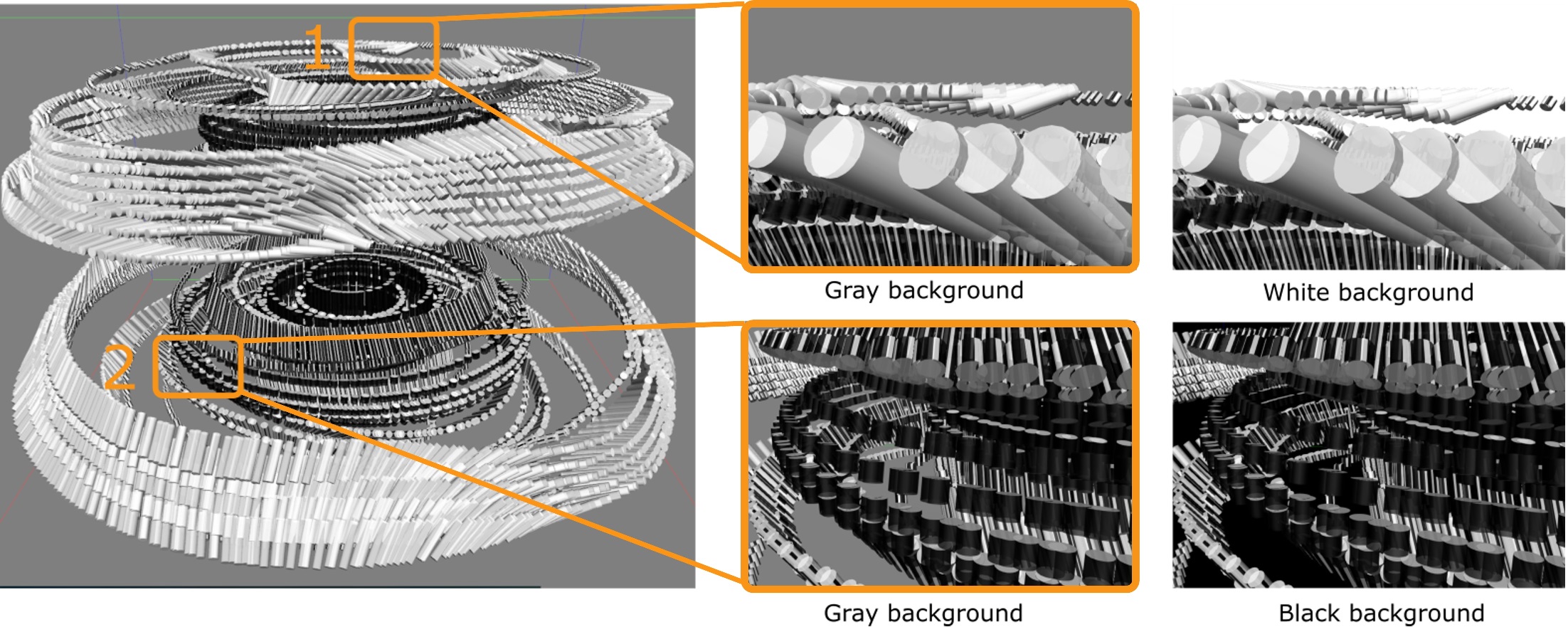}
\caption{Examples using different background colors: gray, white, and black. Figures on the top row are magnified views of region $1$, marked by orange-box on the left image, and the bottom row shows region $2$. With white background, the white cylinders would be washed out (top right image). With black background, the black cylinders would be washed out (bottom right image). In this study, the neutral stimulus-free gray background was chosen. 
}
\label{fig:background}
\end{figure*}

Fig.~\ref{fig:background} shows an example represented by \lengthytexture with gray, white, and black background colors. Gray background color was selected for the experiments. We could observe that both white and black cylinders with \lengthytexture encoding could be displayed more clearly in the gray background (Fig.~\ref{fig:background}, left). 

\section*{B. Visual Mapping for Color and Texture in the \Lengthycolor and \Lengthytexture Pairs}
\label{sec:appendixa}

Fig.~\ref{fig:exp2VisualMapping} shows the visual mapping using color and texture in Experiment II. The horizontal axis represents the exponent range $\in[3, 7]$. We selected those categorical colors from ColorBrewer~\cite{harrower2003colorbrewer}. For texture, the percentage of black is mapped to the exponent-range. Examples with three different  exponent-ranges of 3, 5, and 7 are shown in Fig.~\ref{fig:powerLevelExample}, in which color and texture are used for the visual mapping of study data.

\section*{C. Visual Features and Exponent-Range}
\label{sec:appendixc}

\begin{figure*}[!tp]
	\centering
	\begin{subfigure}[t]{0.9\textwidth}
		\centering
		\includegraphics[width=\textwidth]{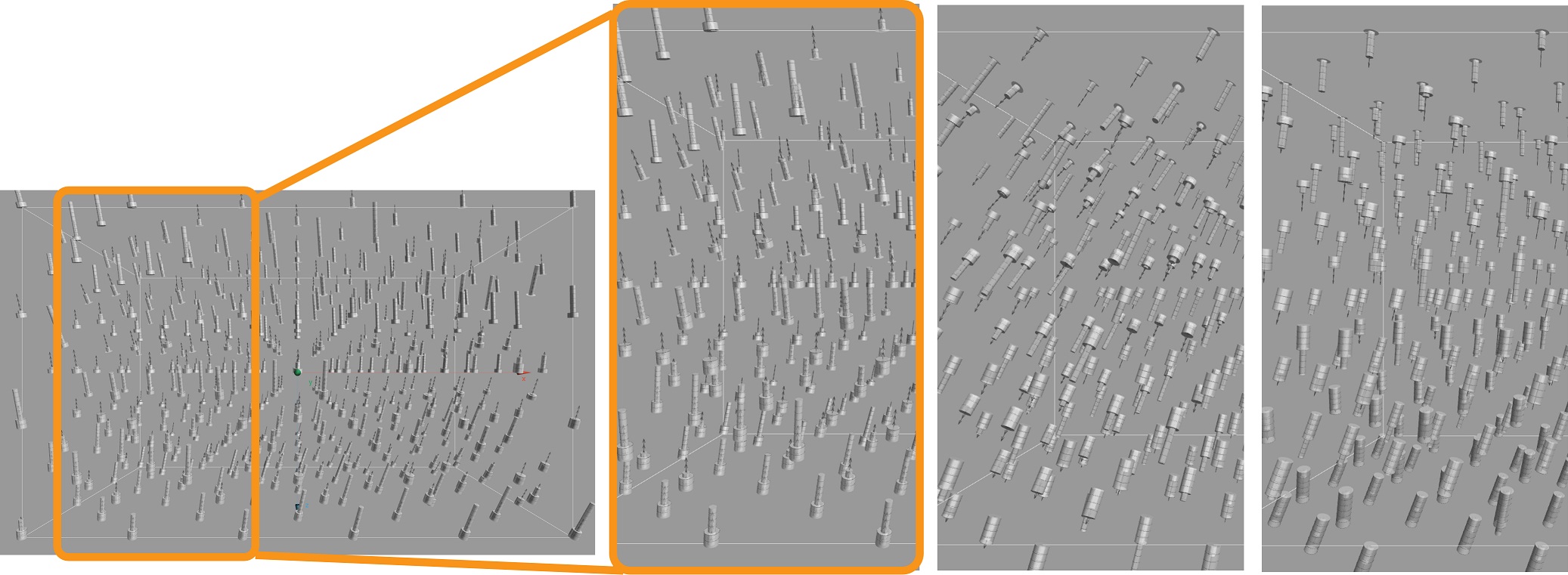}
		\caption{\Lengthylengthy (\textit{splitVectors})}
		\label{fig:lengthPowerExample}
	\end{subfigure}

	\begin{subfigure}[t]{0.9\textwidth}
	\centering
	\includegraphics[width=\textwidth]{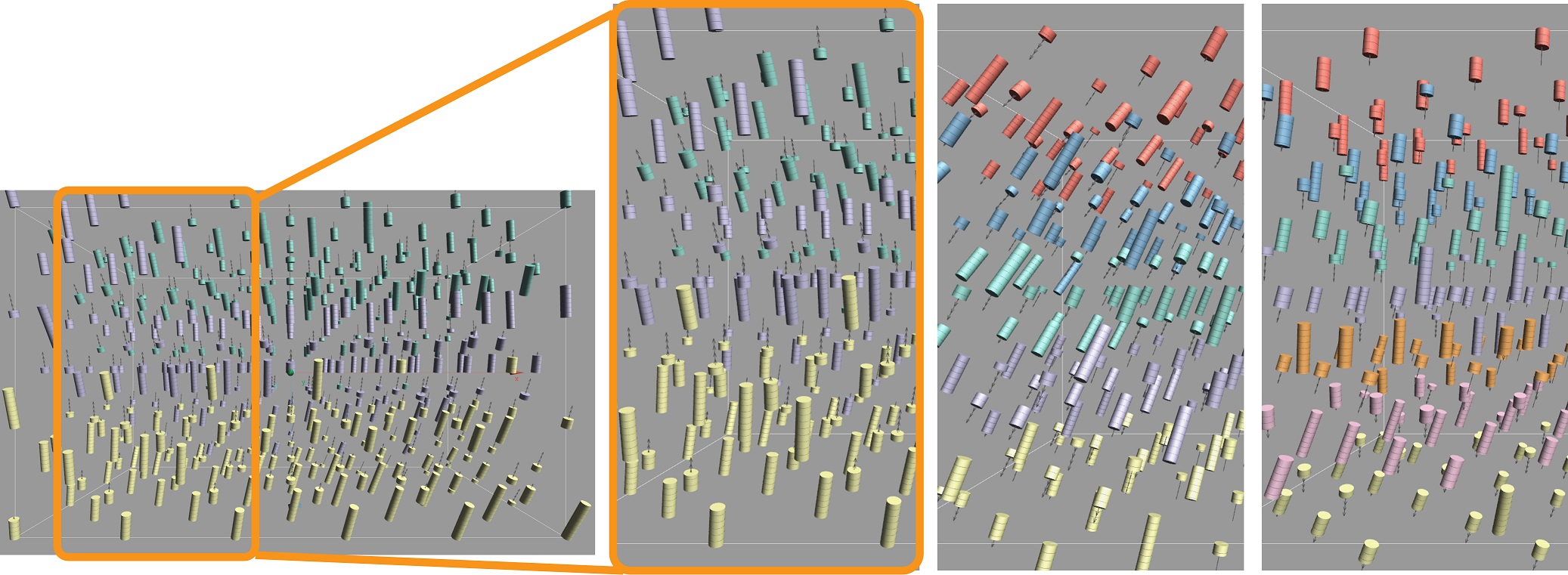}
	\caption{\Lengthycolor}
	\label{fig:colorPowerExample}
	\end{subfigure}

	\begin{subfigure}[t]{0.9\textwidth}
	\centering
	\includegraphics[width=\textwidth]{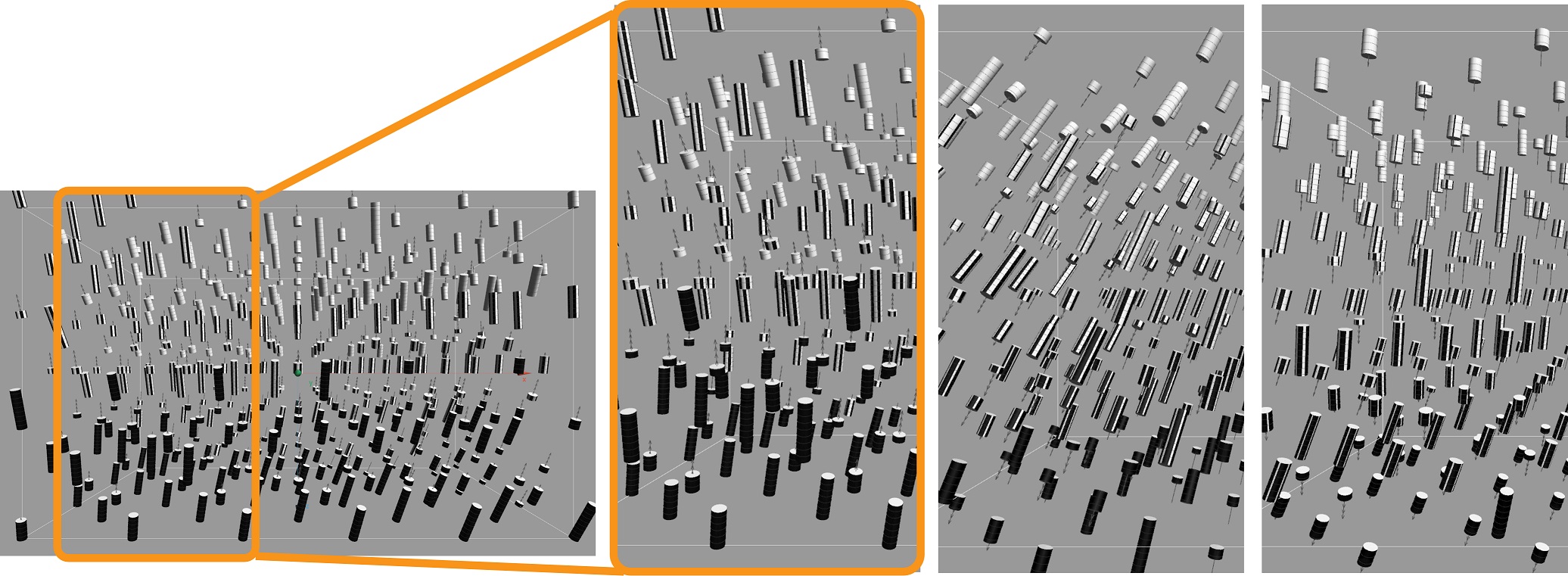}
	\label{fig:texturePowerExample}
	\caption{\Lengthytexture}
	\end{subfigure}
	
	\caption{\changesr{Experiment II: examples of selected exponent ranges of $3$, $5$, and $7$ (from the second left to right). We could see that the pattern of magnitude distribution is more revealing by categorical colors than by texture glyphs. 
Coloring may show more steps with large exponent ranges and also give us a better understanding of data distribution. For example, we could quickly focus on the orange region.}
	}
	\label{fig:powerLevelExample}
\end{figure*}

Fig.~\ref{fig:powerLevelExample} shows examples for visual features and three exponent-ranges of $3$, $5$, and $7$. The figures with the same exponent-range were generated using the same data and different visual features. The dataset used in this figure is for illustration purpose only and does not necessarily reflect all image features used in the vector magnitude experiments.

\section*{D. Spatial Proximity}

\begin{figure*}[!hp]
	\centering
	\begin{subfigure}[t]{0.33\textwidth}
		\centering
		\includegraphics[width=\textwidth]{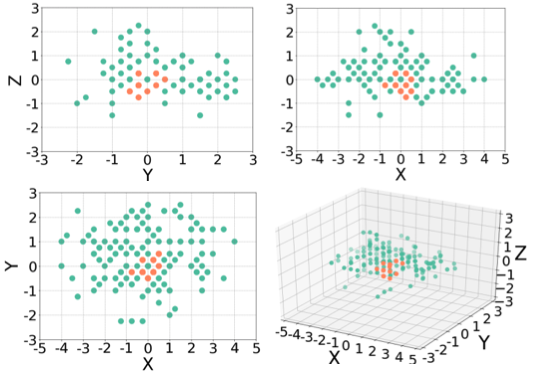}
		\caption{\lengthycolor}
	\end{subfigure}
    \begin{subfigure}[t]{0.33\textwidth}
		\centering
		\includegraphics[width=\textwidth]{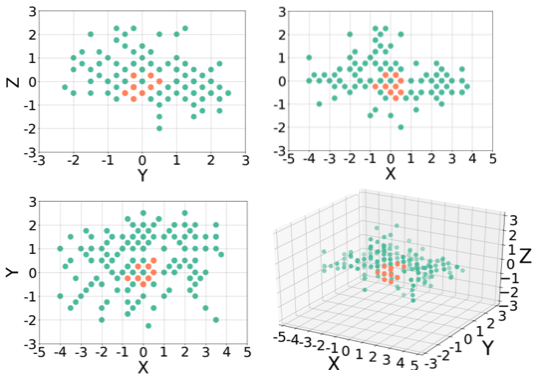}
		\caption{\lengthytexture}
	\end{subfigure}
	\begin{subfigure}[t]{0.33\textwidth}
		\centering
		\includegraphics[width=\textwidth]{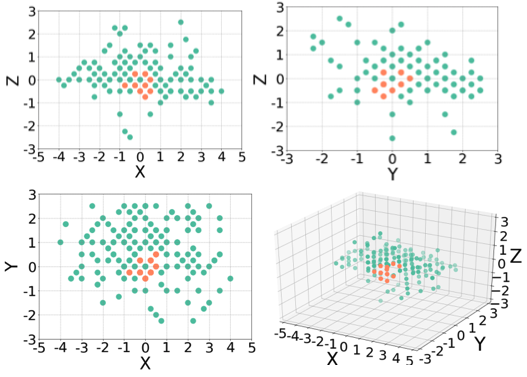}
		\caption{\lengthylengthy}
	\end{subfigure}
\caption{\changesr{Experiment II: Search task. The spatial proximity of the locations of the identified targets, to the ground truth, for all trials in the study. Here the ground truth locations are translated to the origin (0, 0, 0).
This task was time-constrained. among the 810 trials (or 270 trials for each bivariate glyph type), participants completed 262 \lengthycolor, 261 \lengthytexture, and 251 \lengthylengthy trials.
}}
	\label{fig:searchSpatialDis}

\end{figure*}

\begin{figure*}[!hp]
	\centering
	\begin{subfigure}[t]{0.33\textwidth}
		\centering
		\includegraphics[width=\textwidth]{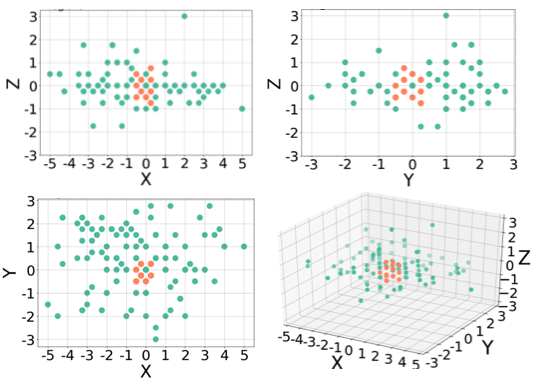}
		\caption{\lengthycolor}
	\end{subfigure}
    \begin{subfigure}[t]{0.33\textwidth}
		\centering
		\includegraphics[width=\textwidth]{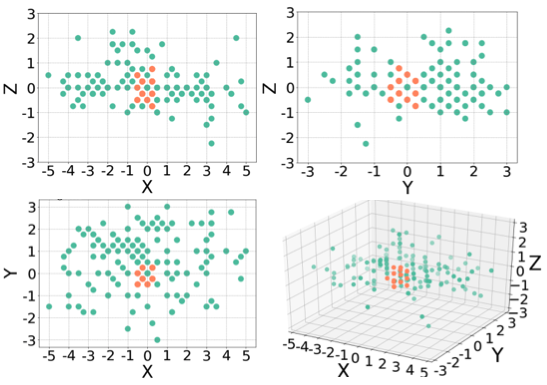}
		\caption{\lengthytexture}
	\end{subfigure}
    \begin{subfigure}[t]{0.33\textwidth}
		\centering
		\includegraphics[width=\textwidth]{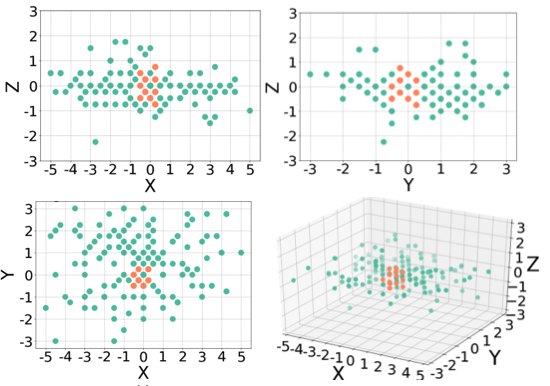}
		\caption{\lengthylengthy}
	\end{subfigure}
	\caption{\changesr{Experiment II: Max task. The spatial proximity of the locations of the identified targets, to the ground truth (centered at the origin (0, 0, 0), for all trials in this task. The yellow dots show 
	the closest points from other-than-target-exponent regions. 
	Here the ground truth locations are translated to the origin (0, 0, 0).
 Among the 810 trials, participants gave an answer to 270 trials for each bivariate glyph type. Among each of these 270, participants completed 269 \lengthycolor, 269 \lengthytexture, and 259 \lengthylengthy trials in total. 
}}
		\label{fig:maxSpatialDis}

\end{figure*}

\changesr{
Figures~\ref{fig:searchSpatialDis} and ~\ref{fig:maxSpatialDis} show spatial distributions of the identified targets (participants' answers) to the correct targets in the search and max tasks in Experiment II.  Here locations of the correct targets are translated to the origin (0, 0, 0). Participants' answers are depicted in green and each dot represents a trial. Dots may overlap. Dots in orange illustrate some of the nearest spins whose exponent values differ from the target (located at the origin). Comparing the distribution of participants' answers and the orange dot locations illustrates one of the key quantum physics data attributes:  quantum physics data are discrete; and spatial proximity is not correlated with the spin magnitude proximity. For complex data like this, 
using the structural features (e.g., from color) in search will help them be more efficient and reduce errors.
}

\end{document}